\documentclass[aps,prc,reprint,showpacs,superscriptaddress]{revtex4-1}

\usepackage{graphicx,color,rotating,pifont}
\usepackage{amsmath,amssymb,bm}
\usepackage{bbm}
\usepackage{ae}
\usepackage{pstricks}
\usepackage{hyperref}

\usepackage{float}
\usepackage{subfigure}
\DeclareMathSymbol{\NS}{\mathord}{AMSb}{"4E}

\newcommand{\ket}[1]{\ensuremath{\,|{#1}\rangle}}
\newcommand{\bra}[1]{\ensuremath{\langle{#1}|}}
\newcommand{\braket}[2]{\ensuremath{\langle{#1}|{#2}\rangle}}
\newcommand{\ketbra}[2]{\ensuremath{|{#1}\rangle\langle{#2}|}}
\newcommand{\matrixe}[3]{\ensuremath{\langle{#1}|\,{#2}\,|{#3}\rangle}}

\newcommand{\diag}{\mathop{\mathrm{diag}}\nolimits}

\newcommand{\expect}[1]{\ensuremath{\langle{#1}\rangle}}

\newcommand{\comm}[2]{\ensuremath{[{#1},{#2}]}}

\newcommand{\op}[1]{\ensuremath{#1}}
\newcommand{\adj}[1]{\ensuremath{{{#1}}^{\dag}}}

\newcommand{\totd}[2]{\ensuremath{ \frac{d {#1}} {d {#2}} }}

\newcommand{\etaO}{\ensuremath{\op{\eta}}}

\newcommand{\HO}{\ensuremath{\op{H}}}

\newcommand{\OO}{\ensuremath{\op{O}}}

\newcommand{\UO}{\ensuremath{\op{U}}}
\newcommand{\VO}{\ensuremath{\op{V}}}

\newcommand{\UUO}{\ensuremath{\adj{\op{U}}}}

\newcommand{\idO}{\ensuremath{\mathbbm{1}}}

\newcommand{\qV}{\ensuremath{\vec{q}}}

\newcommand{\Vlowk}{\ensuremath{V_{\text{low-k}}}}

\newcommand{\EMax}[1][]{\ensuremath{E_{#1\text{max}}}}
\newcommand{\NMax}{\ensuremath{N_{\text{max}}}}
\newcommand{\LMax}{\ensuremath{L_{\text{max}}}}
\newcommand{\eMax}{\ensuremath{e_{\text{max}}}}

\newcommand{\nuc}[2]{\ensuremath{^{#2}\mathrm{#1}}}

\newcommand{\fm}{\ensuremath{\,\text{fm}}}
\newcommand{\fmi}{\ensuremath{\,\text{fm}^{-1}}}

\newcommand{\keV}{\ensuremath{\,\text{keV}}}
\newcommand{\MeV}{\ensuremath{\,\text{MeV}}}

\newcommand{\NNNLO}{N$^3$LO}
\newcommand{\NNLO}{NNLO}

\definecolor{FGViolet}{rgb}{0.61,0.32,0.61}
\definecolor{FGDarkBlue}{rgb}{0,0,0.6}
\definecolor{FGBlue}{rgb}{0,0,0.8}
\definecolor{FGLightBlue}{rgb}{0.2, 0.6, 0.8}
\definecolor{FGGreen}{rgb}{0.2,0.7,0.2}
\definecolor{FGLightGreen}{rgb}{0.4,1,0.4}
\definecolor{FGYellow}{rgb}{1,0.95,0}
\definecolor{FGOrange}{rgb}{0.95,0.5,0.1}
\definecolor{FGRed}{rgb}{0.8,0,0}
\definecolor{FGWhite}{rgb}{1,1,1}
\definecolor{FGLightGray}{rgb}{0.8,0.8,0.8}
\definecolor{FGGray}{rgb}{0.5,0.5,0.5}
\definecolor{FGDarkGray}{rgb}{0.3,0.3,0.3}
\definecolor{FGBlack}{rgb}{0,0,0}

\begin{document}
\title{Singular Value Decomposition and Similarity Renormalization Group Evolution of Nuclear Interactions}

\author{B. Zhu}
\affiliation{Facility for Rare Isotope Beams, Michigan State University,
East Lansing, MI 48824-1321}
\affiliation{Department of Physics \& Astronomy, Michigan State University,
East Lansing, MI 48824-1321}
\email{Corresponding author. E-mail: zhub@frib.msu.edu}
\author{R. Wirth}
\email{E-mail: wirth@frib.msu.edu}
\affiliation{Facility for Rare Isotope Beams, Michigan State University,
East Lansing, MI 48824-1321}
\author{H. Hergert}
\email{E-mail: hergert@frib.msu.edu}
\affiliation{Facility for Rare Isotope Beams, Michigan State University,
East Lansing, MI 48824-1321}
\affiliation{Department of Physics \& Astronomy, Michigan State University,
East Lansing, MI 48824-1321}

\date{\today}

\begin{abstract}
One of the main challenges for \emph{ab initio} nuclear many-body theory in the coming decade is the growth of computational and storage costs as calculations are extended to increasingly heavy, exotic, and structurally complex nuclei. Here, we investigate the factorization of nuclear interactions as a means to address this issue. We perform Singular Value Decompositions of current nucleon-nucleon interactions in partial wave representation and study the dependence of the singular value spectrum on interaction characteristics like regularization schemes and resolution scales. 

Next, we develop and implement the Similarity Renormalization Group (SRG) evolution of the interaction in terms of the relevant singular vectors, and demonstrate that this SVD-SRG approach accurately preserves two-nucleon observables.

We find that low-resolution interactions naturally allow the truncation of the SVD at low rank, and that a small number of relevant components is sufficient to capture the nuclear interaction and perform an accurate SRG evolution, while the Coulomb interaction requires special consideration. The rank is uniform across all partial waves and almost independent of the basis choice in the tested cases. This suggests an interpretation of the relevant singular components as mere representations of a small set of abstract operators that can describe the interaction and its SRG flow.

Following the traditional workflow for nuclear interactions, we discuss how the transformation between the center-of-mass and laboratory frames creates redundant copies of the partial wave components when implemented in matrix representation, and we discuss strategies for mitigation.

Finally, we test the low-rank approximation to the SRG-evolved interactions in many-body calculations using the In-Medium SRG. By including nuclear radii in our analysis, we verify that the implementation of the SRG using the singular vectors of the interaction does not spoil the evolution of other observables.
\end{abstract}

\maketitle

\clearpage

\section{Introduction\label{sec:intro}}
Over the past decade, the reach of \emph{ab initio} nuclear many-body methods across the nuclear landscape has grown dramatically (see, e.g., \cite{Hergert:2020am} and references therein). A new generation of methods that solve the many-body Schr\"odinger equation with controlled approximations have made routine calculations for nuclei up to the tin region possible, and recent progress in the handling of the input three-nucleon interactions \cite{Miyagi:2021ab} will pave the way for applications to even heavier nuclei. 
As the \emph{ab initio} nuclear structure community heads into a new decade, we are facing a number of challenges. The description of doubly open-shell and exotic nuclei require the use of single-particle bases that can naturally account for nuclear deformation as a means to capture strong collective correlations, as well as the coupling to the continuum. Both types of physics typically entail a ten- to hundredfold increase of the basis size compared to existing applications for (semi-)closed shell nuclei. In this way, the computational effort for current ``workhorse'' versions of methods like the In-Medium Similarity Renormalization Group (IMSRG) \cite{Hergert:2016jk,Hergert:2017kx,Stroberg:2019th,Tichai:2021ab}, Coupled Cluster (CC) \cite{Hagen:2014ve,Hagen:2016rb,Duguet:2015ye,Signoracci:2015mz}, self-consistent Green's Function theory (SCGF) \cite{Dickhoff:2004fk,Soma:2020lo} or even finite-order Many-Body Perturbation Theories (MBPT) \cite{Shavitt:2009,Tichai:2018kx,Tichai:2020ft} increases by several orders of magnitude, turning them from problems that are tractable with capacity resources like small computing clusters to problems that require (or exceed) the capabilities of the largest available supercomputers. 

Another important effort is the proper quantification of errors incurred by the approximations inherent to these methods. This requires access to (at least) the next order of truncation so that the convergence (or lack thereof) towards the exact result can be established, and this implies additional order-of-magnitude increases of the computational cost even for closed-shell nuclei. For example, the commonly used IMSRG(2) truncation has a (naive) computational scaling of $\mathcal{O}(N^6)$ with the single-particle basis size $N$, which increases to $\mathcal{O}(N^7)$ or $\mathcal{O}(N^9)$ for approximate or complete versions of the next-order IMSRG(3) truncation, respectively \cite{Morris:2016xp,Heinz:2021mk}.

Last but not least, the propagation of the theoretical uncertainties of the input interactions and many-body methods to the final results for observables, and the related task of studying the sensitivity of observables to the theoretical parameters both rely on the capability to perform large ensembles of many-body calculations. There have been recent breakthroughs in the use of emulators to tackle this problem \cite{Konig:2020fn,Ekstrom:2019tw,Furnstahl:2020yf}, but the construction of accurate surrogate models still requires a substantial amount of expensive many-body calculations to provide training data. 

A major factor in the computational and storage costs of modern many-body methods is an incompatibility between the representations that are best suited for the input interactions and the treatment of the many-body system, respectively. Our starting two- and three-nucleon Hamiltonians consists of a few tens of operator structures that are built from the spins, isospins and Jacobi momentum or position vectors of the interacting particles, reflecting the fundamental symmetries of space(time) as well as the intrinsic symmetries of the strong interaction (see, e.g., \cite{Wiringa:1995or,Epelbaum:2009ve}). 

Ideally, these operators would be represented in states that directly incorporate the same symmetries, but unfortunately, the construction of such bases is only feasible in few-body systems due to the high cost of properly implementing their antisymmetry under permutations (see, e.g., \cite{Nogga:2001no,Navratil:2000hf,Nogga:2006yg,Barnea:1997vn,Barnea:2004cr}). 
For many-body systems, one therefore uses a basis of Slater determinants, which are antisymmetrized by construction. Of course, the drawback of these states is that they describe independent-particle systems and are therefore ill-suited for capturing the correlations that are induced by nuclear interactions. 

Consequently, a typical workflow increases the number of interaction matrix elements by several orders of magnitude as they are transformed from their initial representation in Jacobi coordinates to the laboratory-frame representation used by the many-body method. For instance, two nucleon force matrix elements grow from a few MB to hundreds of MBs. For three-nucleon forces, the growth is even worse,       and typically only manageable by imposing severe truncations on the laboratory-frame matrix elements \cite{Roth:2014fk,Binder:2014fk,Miyagi:2021ab}. 

The transformation between the center-of-mass and laboratory frames consists of the construction of a tensor product of the interaction in relative partial waves with an identity operator acting on the center-of-mass wave function of the particles, and a subsequent basis change (see Sec.\ref{sec:talmi}). At no point do we introduce new, physically relevant information to the matrix elements, hence the aforementioned growth of the matrix element sets is entirely owed to the inefficiencies of the laboratory-frame representation. This observation strongly suggests that it would be fruitful to perform a principal component analysis (PCA) of the interaction matrix elements in order to recover the essential components in each representation. Naively, one could expect their number to be close to the number of operator structures and other pieces of physical information in the nuclear interactions,  e.g., LECs, cutoffs, and characteristics of the radial or momentum dependencies. However, we wish to maintain a connection to the aforementioned working bases, so that the principal components of the interaction can be integrated efficiently into existing nuclear many-body methods, hence some amount of inefficiency is unavoidable. 

In this work, we initiate a larger program for the PCA of modern nuclear interactions by focusing on nucleon-nucleon ($NN$) interactions, specifically. Our central tool is the singular value decomposition (SVD). The SVD allows us to uncover the low-rank structure of the $NN$ interaction, which will be carried forward through a free-space SRG evolution that dials the resolution scale (Sec. \ref{sec:srgsvd}) and the transformation to the laboratory frame (Sec. \ref{sec:talmi}). We show that these procedures can be implemented efficiently and accurately using the factors obtained by the low-rank decomposition of the interaction (Sec. \ref{sec:test}), setting the stage for applying the same techniques to three-nucleon forces. 

In Sec. \ref{sec:manybody}, we test the rank-reduced interactions --- and associated SRG transformations --- in IMSRG(2) calculations of energies and radii. At present, we reconstruct the laboratory-frame matrix elements from the factors of the interaction, but we eventually intend to exploit the factorized structure to reduce the computational cost of many-body methods for medium-mass and heavy nuclei, along the lines of \cite{Tichai:2019xz} as well as successful application of factorization methods in quantum chemistry \cite{Hohenstein:2012xq,Hohenstein:2012ez,Parrish:2012os,Schutski:2017xr,Parrish:2019zg,Hohenstein:2019jx}. This could provide a means for addressing the challenges discussed earlier in this section, and enable the use of efficient rank-reduced \emph{ab initio} calculations in the day-to-day analysis of experimental data.

\section{Similarity Renormalization Group Evolution of Factorized Interactions}
\label{sec:srgsvd}

In this section, we briefly recapitulate the essential aspects of the SVD and the SRG evolution before merging the two techniques. We introduce some terminology along the way.

\subsection{Singular Value Decomposition}

The SVD can be understood as a generalization of the eigenvalue decomposition to rectangular and non-normal\footnote{A matrix $M$ is \emph{normal} if and only if it commutes with its Hermitian conjugate, $\comm{M}{M^\dag} = 0$. Examples are Hermitian and unitary matrices.} matrices. A general complex $m\times n$ matrix $M$ can be uniquely decomposed as (see, e.g., \cite{Golub:2013le})
\begin{equation} \label{eq:svd}
  M = U \Sigma V^\dag
\end{equation}
where $\UO\in\mathbb{C}^{m\times m}$ and $\VO\in\mathbb{C}^{n\times n}$ are both unitary, and 
\begin{equation}
  \Sigma = \diag(\sigma_1, \ldots, \sigma_R, 0, \ldots )\in\mathbb{R}^{m\times n} 
\end{equation}
is a diagonal matrix whose entries are non-negative in descending order. The number of non-zero singular values defines the \emph{rank} $R\leq \min(m,n)$ of the matrix $M$. 

A \emph{truncated} SVD is obtained by approximating
\begin{equation} \label{eq:svd_t}
  M \approx \sum_{i=1}^r u_i \sigma_i v^\dag_i\,,
\end{equation}
where $r < R$ and $u_i, v_i$ are the \emph{singular vectors}, i.e., the columns of the matrices
$U$ and $V$.

\subsection{Similarity Renormalization Group}
\label{sec:srg}

The (free-space) SRG is a continuous unitary transformation that is designed to decouple the low and high-momentum components of the Hamiltonian and other observables of interest (see, e.g., \cite{Bogner:2010pq,Tropiano:2020ad}). The transformation, or SRG flow, is parameterized with a continuous flow parameter $s$, and implemented through the operator flow equation
\begin{equation} \label{eq:srgflow}
  \totd{}{s} \HO(s) = \comm{\etaO(s)}{\HO(s)}\,,  
\end{equation}
starting from the initial Hamiltonian $H(s=0)$. 

The SRG framework gives us a great degree of freedom in selecting an ansatz for the anti-Hermitian generator $\eta(s)$. Here, we will use the standard ansatz for performing momentum decoupling in nuclear interactions (see \cite{Bogner:2010pq,Hebeler:2020ex} and references therein):
\begin{equation} \label{eq:def_eta}
  \etaO(s) \equiv \comm{T}{H(s)}\,, 
\end{equation}
where $T$ is the relative (or intrinsic) kinetic energy. Note that this implies that the kinetic energy remains \emph{constant} throughout the flow, hence all $s$ dependent contributions from evolving $T$ are absorbed into the interaction $V(s)$: 
\begin{equation} \label{eq:h_partition}
  H(s) = T + \delta T(s) + \overline{V}(s) \equiv T + V(s)\,,
\end{equation}
with
\begin{equation}
  \delta T(0) = 0\,, \qquad \overline{V}(0) =  V(0)\,.
\end{equation}
This partitioning will become relevant in the subsequent discussion. Plugging Eqs. \eqref{eq:def_eta} and \eqref{eq:h_partition} into the flow equation \eqref{eq:srgflow} and using $dT/ds=0$, we obtain the following flow equation for $V(s)$:
\begin{align}
  \totd{}{s}V(s)  &= \comm{\eta(s)}{T} + \comm{\eta(s)}{V(s)}\notag\\
                  &= \comm{\comm{T}{V(s)}}{T} + \comm{\comm{T}{V(s)}}{V(s)}\,.\label{eq:srg_interaction}
\end{align}
This equation is conveniently implemented in momentum space, where $T$ will be diagonal \cite{Bogner:2010pq}.

In principle, general observables can be computed by evolving them alongside the Hamiltonian according to
\begin{equation}
  \totd{}{s}O(s) = \comm{\eta(s)}{O(s)}\,.
\end{equation}
In the free-space SRG, the ensuing growth of the system of flow equations can be avoided in several ways: Since the initial and final Hamiltonian matrices are available, one can solve their respective eigenvalue problems and directly construct $U(s)$ as \cite{Anderson:2010br,Schuster:2014oq,Gysbers:2019df}
\begin{equation}
  U(s) = \sum_n\ket{\psi_n(s)}\bra{\psi_n(0)}\,.
\end{equation}
In fact, any complete basis could be used to express $U(s)$ in this way if all initial and evolved basis vectors are readily available.
One can also determine $U(s)$ directly by solving
\begin{equation}\label{eq:srg_U}
  \totd{}{s}U(s) = \eta(s) U(s)\,,\quad U(s=0) = \idO\,.
\end{equation}
A third alternative is the use of the Magnus expansion, although this method is potentially susceptible to convergence issues \cite{Tropiano:2020ad}.

For future use, we introduce the \emph{resolution scale} of SRG-evolved interactions \cite{Bogner:2010pq} through the definition
\begin{equation}
  \lambda \equiv s^{-1/4}\,,
\end{equation}
which has the dimensions of a momentum if we work with the generator \eqref{eq:def_eta}. It can be understood as a smooth regulator on the momentum transfer between incoming and outgoing states, e.g., 
\begin{equation}
  Q = |\qV_\text{out} - \qV_\text{in}| \lesssim \lambda\label{eq:srg_momentum_transfer}
\end{equation}
in the two-body system, which suggests that the matrix representation of the interaction in momentum space is a (slightly diffuse) band with width $\lambda$ \cite{Bogner:2010pq}.

\subsection{SRG Evolution of SVD Factors}

Let us now assume that we have performed an SVD decomposition of an initial operator $O(0)$,
and consider its SRG evolution. Writing
\begin{equation}\label{eq:svd_O0}
    \OO(0) = \sum_i |u_i(0)\rangle\, \sigma_i\, \langle v_i(0)| \,,
\end{equation}
we see that the evolved operator can be written as
\begin{align}
    \OO(s) &= \sum_i \UO(s)|u_i(0)\rangle \,\sigma_i\, \langle v_i(0)|\UUO(s)\,,\notag\\
           &\equiv \sum_i |u_i(s)\rangle \,\sigma_i\, \langle v_i(s)|\label{eq:svd_Os}
\end{align}
where we have used that singular values are invariant under unitary evolutions, 
and defined the evolved singular vectors $\ket{u_i(s)}$ and $\ket{v_i(s)}$. We can
immediately use Eq.~\eqref{eq:srg_U} to obtain flow equations for these states:
\begin{subequations}\label{eq:svd_flow}
  \begin{align}
      \totd{}{s} |u_i(s)\rangle = \etaO(s)|u_i(s)\rangle\,, \\
      \totd{}{s} |v_i(s)\rangle = \etaO(s)|v_i(s)\rangle\,.
  \end{align}
\end{subequations}

\begin{figure}[t]
  \includegraphics[width=0.48\textwidth]{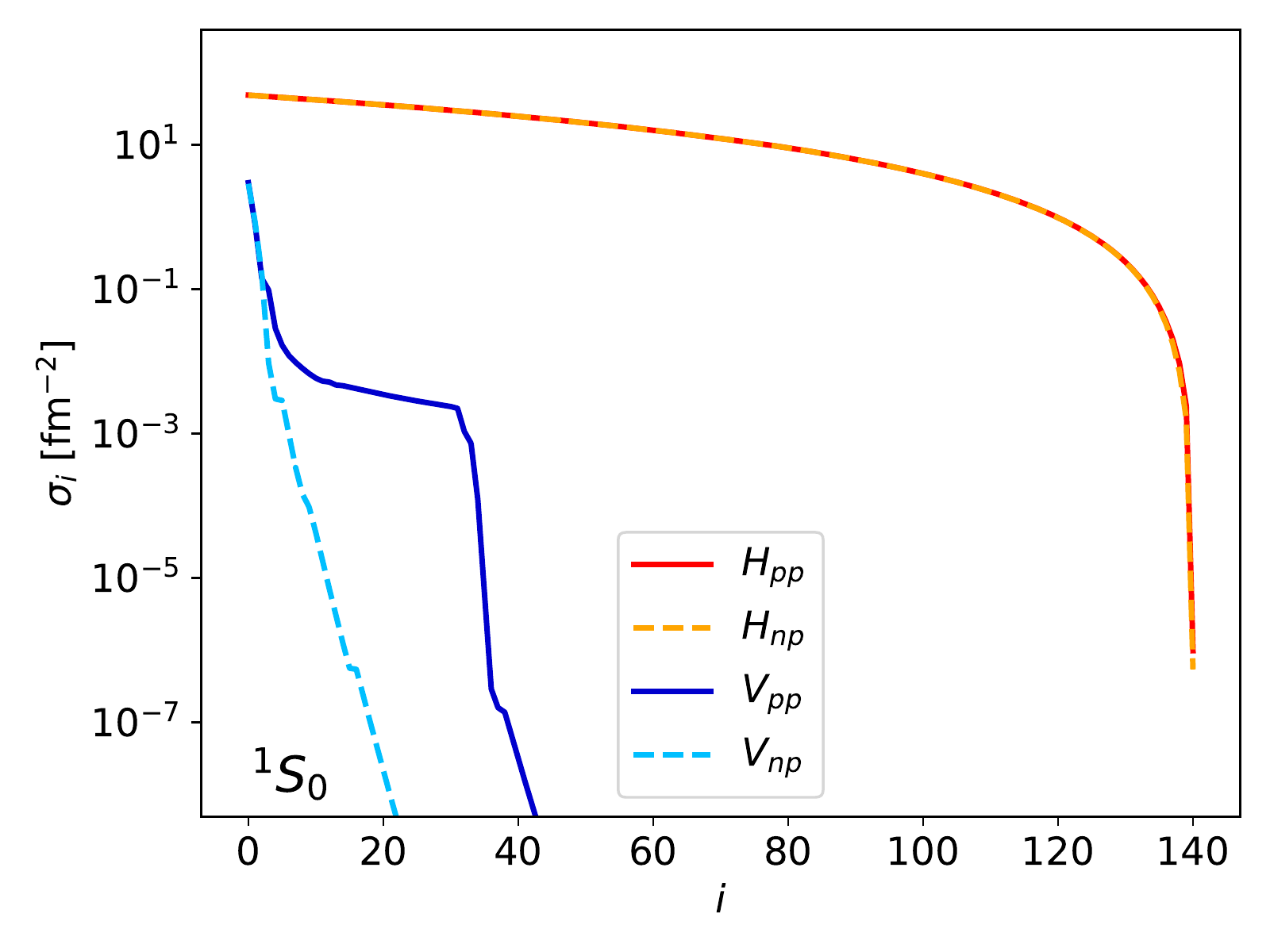}
  \vspace{-20pt}
  \caption{Singular value spectra of $H$ and $V$ in the proton-proton (solid curves) and neutron-proton  ${}^1S_0$ partial waves (dashed curves), for a chiral \NNNLO{} two-nucleon interaction (EM, cutoff $\Lambda=500\,\MeV/c$, see text and Ref.~\cite{Entem:2003th}). The units of the singular vectors result from adopting scattering units $\hbar c/mc^2 = 1$ as well as momentum-space discretization discussed in Sec.~\ref{sec:conventions}. }
  \label{fig:sv_decay}
\end{figure}

If we attempt to implement this form of the SRG flow for the Hamiltonian, we encounter a major issue: Due to the presence of the kinetic energy, which is unbounded from above, there is no natural point at which we can truncate the SVD of $H(0)$, as illustrated in Fig.~\ref{fig:sv_decay}. Thus, we would not gain any numerical advantage from implementing Eqs.~\eqref{eq:svd_flow} over Eqs.~\eqref{eq:srgflow} or \eqref{eq:srg_U}. The solution to this problem is to consider the SVD of the evolving part of the Hamiltonian, namely the interaction $V(s)$ as defined in the partitioning \eqref{eq:h_partition}. 

At $s=0$, the SVD of the interaction is given by
\begin{equation}
  V(0) = \sum_{ij} \ket{u_i(0)}\, \sigma_{ij}\, \bra{v_j(0)}\,, \quad \sigma_{ij} = \sigma_i\delta_{ij}\,,
\end{equation}
where we have used that the interaction will be represented as a square Hermitian matrix. 
As an example, Fig.~\ref{fig:sv_decay} shows the singular value spectrum of the proton-proton and neutron-proton ${}^1S_0$ partial waves of a realistic chiral \NNNLO{} interaction \cite{Entem:2003th}, which reveals the interaction's low-rank structure. As discussed in more detail below, the presence of the Coulomb interaction increases the rank of the interaction in the $pp$ channels, but not to the point where truncations would become unfeasible. A crucial observation is that the singular values of the interaction decay exponentially, while the kinetic term in the Hamiltonian grows only quadratically. This means that the singular value spectrum of the generator \eqref{eq:def_eta} will also decay exponentially, and the SRG evolution via Eq.~\eqref{eq:srg_interaction} cannot spoil the low-rank structure of the interaction, barring truncation artifacts.

For $s>0$, we have
\begin{equation}
  V(s) = \sum_{ij} \ket{u_i(s)}\, \left(\sigma_{ij} + \delta \sigma_{ij}(s)\right)\,\bra{v_{j}(s)}\,.\label{eq:svd_Vs}
\end{equation}
where $\sigma_{ij}$, which contains the singular values of the \emph{initial} interaction, remains constant under unitary evolution, while $\delta \sigma_{ij}(s)$ contains the contributions from the absorbed $s$-dependent part of the kinetic energy.

We can derive a flow equation for $\delta \sigma_{ij}(s)$ by considering the left- and right-hand sides of Eq.~\eqref{eq:srg_interaction}. Plugging Eq.~\eqref{eq:svd_Vs} into the LHS and using the flow equations \eqref{eq:svd_flow}, we obtain
\begin{align}
  \totd{}{s} V(s) &
  =\comm{\eta(s)}{V(s)} + \sum_{ij}\ket{u_i(s)}\left(\totd{}{s}\delta\sigma_{ij}(s)\right)\bra{v_j(s)}\label{eq:svd_Vs_lhs}\,.
\end{align}
Comparing with the RHS and using the orthogonality of the singular vectors, we obtain
\begin{equation}
  \totd{}{s}\delta\sigma_{ij}(s) = \matrixe{u_i(s)}{\comm{\eta(s)}{T}}{v_j(s)}\,.
\end{equation}
In principle, this flow equation would have to be solved alongside with Eq.~\eqref{eq:svd_flow}, but it turns out that it can be integrated analytically. Expanding the right-hand side and using Eq.~\eqref{eq:svd_flow} to switch to derivatives of the singular vectors, we quickly obtain the closed solution
\begin{equation}\label{eq:delta_sigma}
  \delta\sigma_{ij}(s) = \matrixe{u_i(0)}{T}{v_j(0)} - \matrixe{u_i(s)}{T}{v_j(s)}\,,
\end{equation}
i.e., we merely need to compute the matrix elements of the kinetic energy in the representation spanned by the initial and final singular vectors. As a consistency check, we note that the first term in this equation can be written as
\begin{align}
  \matrixe{u_i(0)}{T}{v_j(0)} &= \matrixe{u_i(0)}{\UUO(s)\UO(s) T \UUO(s) \UO(s)}{v_j(0)} \notag\\
    &= \matrixe{u_i(s)}{T +\delta T(s) }{v_j(s)}\,,
\end{align}
so Eq.~\eqref{eq:delta_sigma} is the representation of the induced, $s$-dependent part of the kinetic energy in the (possibly truncated) basis of \emph{evolved} singular vectors: 
\begin{equation}\label{eq:delta_T}
  \delta\sigma_{ij}(s) = \matrixe{u_i(s)}{\delta T(s)}{v_j(s)}\,.
\end{equation}

As discussed in Sec.~\ref{sec:srg}, we can easily construct the SRG transformation $U(s)$ if we have access to the initial and evolved versions of a complete basis set. Both the $|u_i(s)\rangle$ and $|v_i(s)\rangle$ qualify as such bases, hence we have
\begin{subequations}\label{eq:unitarySVD}
  \begin{align}
      \UO^{(L)}(s) &= \sum_i |u_i(s)\rangle\langle u_i(0)|\,,\\
      \UO^{(R)}(s) &= \sum_i\ket{v_i(s)}\bra{v_i(0)}\,,
  \end{align}
\end{subequations}
if we need to distinguish left and right unitary evolutions, and we can construct approximate unitaries from the truncated basis sets. For Hermitian matrices, the left and right singular vectors merely differ by a trivial phase factor, and the two unitaries are identical. Using the exact or approximate $U(s)$, we can transform arbitrary observables $O$ according to
\begin{equation}
    \OO(s)=\UO(s)\OO(0)\UUO(s)\,.
\end{equation}

\section{Applications in the Two-Nucleon System}
\label{sec:test}
We are now ready to analyze SVD-factorized $NN$ interactions and their SRG evolution. In the present section, we will focus on the two-nucleon system, which allows us to work with partial waves in relative-momentum and HO representations. 

\subsection{Momentum Space Discretization and Other Conventions}
\label{sec:conventions}
A commonly used basis for exchanging momentum space matrix elements and performing their SRG evolution is built from states that satisfy the completeness and orthogonality relations
\begin{equation}
  \idO = \sum_{lsjTM_T}\int_{0}^\infty dq\,q^2\, \ketbra{qlsjmTM_T}{qlsjmTM_T}
\end{equation}
and
\begin{align}
  &\braket{qlsjmTM_T}{q'l's'j'm'T'M_T'} = \notag\\
   &\qquad\qquad\frac{\delta(q-q')}{qq'}\delta_{ll'}\delta_{ss'}\delta_{jj'}\delta_{TT'}\delta_{M_TM_T'}\,,
\end{align}
where $s, l$ and $j$ refer to the spin, orbital and total angular momentum of the interacting nucleon pair, and $T$ is the coupled isospin. The quantum numbers $m$ and $M_T$ are the projections associated with $j$ and $T$, respectively. Suppressing the angular momentum and isospin quantum numbers for brevity, the discretized versions of these relations are
\begin{equation}
  \idO = \sum_i\,w_i q_i^2\, \ketbra{q_i}{q_i}
\end{equation}
and 
\begin{equation}
  \braket{q_i}{q_j} = \frac{\delta_{ij}}{\sqrt{w_i w_j} q_iq_j}\,,
\end{equation}
where $w_i$ are the weight factors of our chosen momentum mesh. If we apply the SVD to a matrix represented in this basis, the singular values will acquire an undesirable dependence on the weights $w_i$ because the basis states are not normalized to 1. Thus, we choose to introduce the rescaled states
\begin{equation}
  \ket{\overline{q}_i} = \sqrt{w_i}q_i \ket{q_i}\,,
\end{equation}
which satisfy 
\begin{equation}
  \braket{\overline{q}_i}{\overline{q}_j} = \delta_{ij}\,.
\end{equation}
The matrix elements of the interaction in the rescaled and original bases are related by
\begin{equation}
  \matrixe{\overline{q}_i}{V}{\overline{q}_j} = \sqrt{w_iw_j}q_iq_j\matrixe{q_i}{V}{q_j}\,.
\end{equation}

An immediate benefit of the basis change is that the discretized integration measure --- i.e., the weights and $q_i^2$ factors --- no longer appear explicitly in our working equations, e.g., the flow equations \eqref{eq:svd_flow} \cite{Bogner:2010pq}. Furthermore, it makes it easier to relate the truncated SVD across different mesh and basis choices, which might make it easier in the future to interpret its components as mere representations of the relevant \emph{operators} in the interaction (cf.~Sec.~\ref{sec:ho}). To compare with the existing literature on the SRG evolution of matrix elements (see, e.g., \cite{Bogner:2010pq,Hebeler:2020ex} and references therein), we will simply revert to the original basis. 

\begin{figure}[t]
  \includegraphics[width=0.48\textwidth]{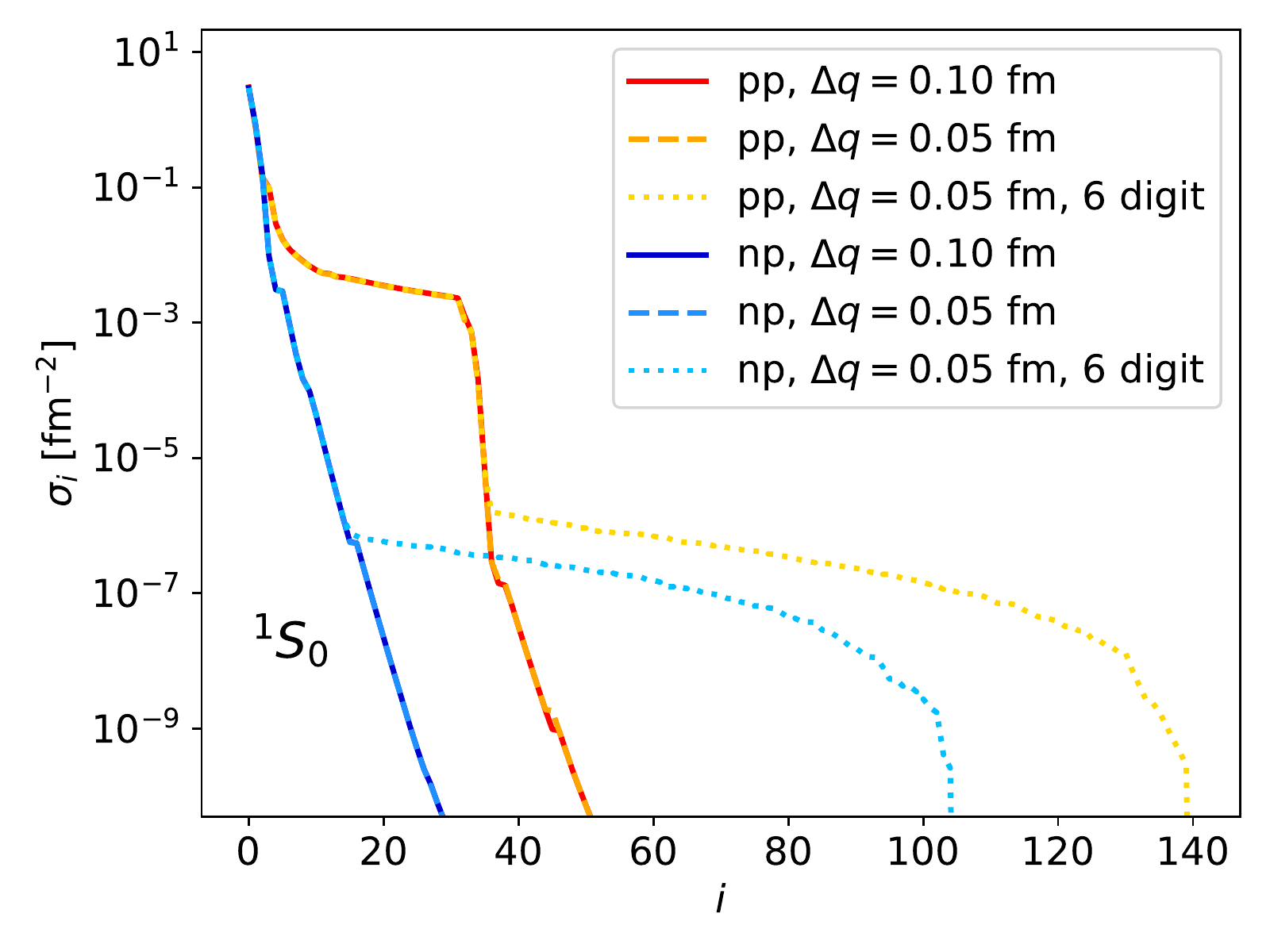}
  \vspace{-20pt}
  \caption{Singular value spectra of $V$ in the proton-proton and neutron-proton  ${}^1S_0$ partial waves, for the EM interaction. The solid and dashed curves are obtained on equidistant meshes with the same $q_\text{Max} = 7\fmi$, but different spacings $\Delta q$. The dotted curve illustrates that the tail of the singular value spectrum is related to the precision of the input data, which is reduced from 10 digits for the other data sets to 6 digits after the decimal. All conventions are the same as in Fig.~\ref{fig:sv_decay} (also cf.~Sec.~\ref{sec:conventions}).}
  \label{fig:sv_mesh}
\end{figure}

In Figure \ref{fig:sv_mesh}, we demonstrate the desired independence of the singular values on the chosen mesh for the proton-proton and neutron-proton partial waves of the nuclear interaction. The calculations are based on the chiral \NNNLO{} interaction with cutoff $\Lambda=500\,\MeV/c$ by Entem and Machleidt \cite{Entem:2003th}, which we will refer to as the EM interaction in the following. The singular value spectra for two equidistant meshes with the same maximum momentum $q_\text{Max}=7\,\fmi$ but different spacings $\Delta q =0.05\,\fmi$ and $0.1\,\fmi$ are practically identical for $\sigma_i \geq 10^{-10}\,\fm^{-2}$. The precision with which the input matrix elements are stored has an impact on the extension of the singular value spectrum: If we reduce the number of decimal digits from 10 (for the solid and dashed curves) to 6, we obtain an elongated tail of unphysical singular values of order $\sigma_i\approx 10^{-6}\,\fm^{-2}$. In practice, of course, we would not consider these contributions to the interaction anyway, and introduce a threshold like $\sigma_t\approx 10^{-3}\,\fm^{-2}$ which emerges naturally for the proton-proton ${}^1S_0$ partial wave (see Sec.~\ref{sec:sv_spectra}).

\subsection{Singular Value Spectra of Nucleon-Nucleon Interactions}
\label{sec:sv_spectra}
With the technical discussion out of the way, we can now analyze the SVDs of $NN$ interactions like the EM \cite{Entem:2003th} and Argonne V18 \cite{Wiringa:1995or} potentials. Both represent nucleon-nucleon scattering data with high accuracy, while having quite different characteristics: As mentioned before, EM is derived within the framework of chiral Effective Field Theory (EFT) (see, e.g., \cite{Epelbaum:2020rr} and references therein) and as a result, has a moderate initial cutoff. Due to the adopted regularization scheme, the interaction is nonlocal, and best represented in momentum space. In contrast, AV18 is designed for applications in coordinate-space Quantum Monte Carlo (QMC) calculations \cite{Gandolfi:2020fv,Lynn:2019dw}, and therefore as local as possible by construction. This results in a strong, repulsive core in the interaction that makes its use problematic in many-body methods that rely on basis expansions. Both EM and AV18 have been superseded by younger cousins from their respective development tracks, but they still represent a useful case study because they contain all essential features that are also present in more recently developed interactions (see, e.g., the recent reviews \cite{Entem:2020oq,Epelbaum:2020rr,Piarulli:2020dp}).

In Figs.~\ref{fig:sv_decay} and \ref{fig:sv_mesh}, we have already seen the ${}^1S_0$ partial waves of the EM interaction. The singular value spectra clearly indicate that the interaction is inherently of low rank in these channels. In the  proton-proton (pp) channel, we observe a ``kink'' that suggests a threshold value of $\sigma_t = 10^{-3}\,\fm^{-2}$ as a natural threshold value for truncating the SVD. This $\sigma_t$ is about three to four orders of magnitude smaller than the largest singular values in the proton-proton and neutron-proton ($np$) channels across all partial waves up to $j=9$, which are $\sigma_\text{max}(pp) \approx \sigma_\text{max}(np) \approx 3\,\fm^{-2}$. 

\begin{figure}
  \includegraphics[width=0.48\textwidth]{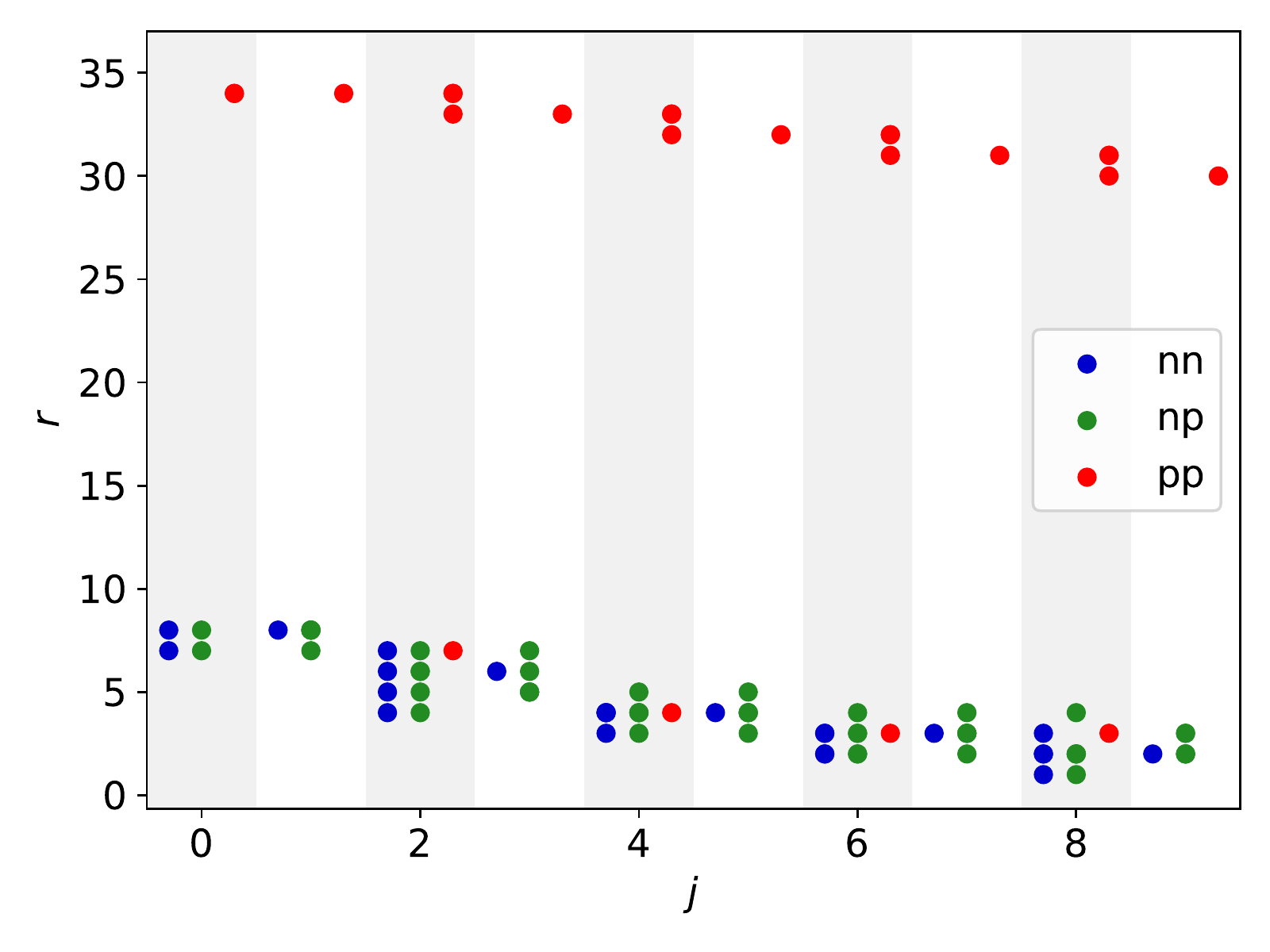}
  \vspace{-20pt}
  \caption{Truncated SVD rank $r$ of partial waves up to $j=9$, using a singular value threshold $\sigma_t = 10^{-3} \fm^{-1}$ (cf. Fig.~\ref{fig:sv_decay} and Sec.~\ref{sec:conventions}). The maximal singular values for the interaction channels are $\{\sigma_\text{max}(nn), \sigma_\text{max}(np), \sigma_\text{max}(pp)\} = \{3.00 \fmi, 2.97 \fmi, 3.06 \fmi\}$, respectively.  }
  \label{fig:num_of_comp}
\end{figure}

In the pp channel, about 35 components of the interaction have singular values above the threshold. In contrast, only about 10 components are above threshold in the $np$ channel. The cause of this difference is readily apparent: It is the inclusion of the 
Coulomb interaction, $V_C$ in the initial matrix elements. This interaction is long-ranged and has no inherent scale other than a hard cutoff at $15\,\fm$ that we impose during the numerical computation of matrix elements. This cutoff is well beyond the size of a nucleus, so this treatment is safe for nuclear structure calculations, although more care is required in scattering applications and reaction theory. Without the cutoff or some other regulator, $(V_C)_{ij} \sim (q_i - q_j')^{-2}$ would be represented by a matrix with a divergence on the diagonal, which would completely spoil our capability to truncate the SVD in the pp channels. 

The relative simplicity of $V_C$ compared to nuclear interactions could make it possible to include it efficiently by means other than a truncated SVD in a chosen configuration space. In the context of the present work, we cannot completely separate the treatment of $V_C$ because it couples to the other terms in the Hamiltonian during the SRG evolution, even if $V_C$ only evolves weakly \cite{Bogner:2010pq}. One could attempt to construct an SRG evolution from $T$ and the nuclear interaction $V_N$ and apply it to $V_C$ after the fact, just like other observables, but this leads to effects on the order of several percent on the ground-state energies of medium-mass nuclei, which is comparable or greater than other theoretical uncertainties for these quantities. This subject deserves more detailed exploration in the future.

Moving to higher partial waves, a consistent picture emerges: As shown in Fig.~\ref{fig:num_of_comp}, the number of singular values above the threshold $\sigma_t=10^{-3}\,\fm^{-2}$ is between 5 and 10 for $np$ and neutron-neutron partial waves (nn) up to $j=9$, while $pp$ partial waves contain about 35 singular values above the threshold due to the presence of the Coulomb interaction. For off-diagonal pp partial waves like ${}^{3}P_2-{}^{3}F_2$, we note that the number of components is comparable to that in the $np$ and nn channels. Since the Coulomb interaction cannot contribute to partial waves with $l\neq l'$, this is further evidence that $V_C$ is the cause for the increased number of relevant components in the $l=l'$ partial waves. 

\begin{figure}
\setlength{\unitlength}{\columnwidth}
\begin{picture}(2.0000,1.18000)
  \put(-0.0200,0.5800){\includegraphics[width=\unitlength,viewport=0 10 456 282,clip]{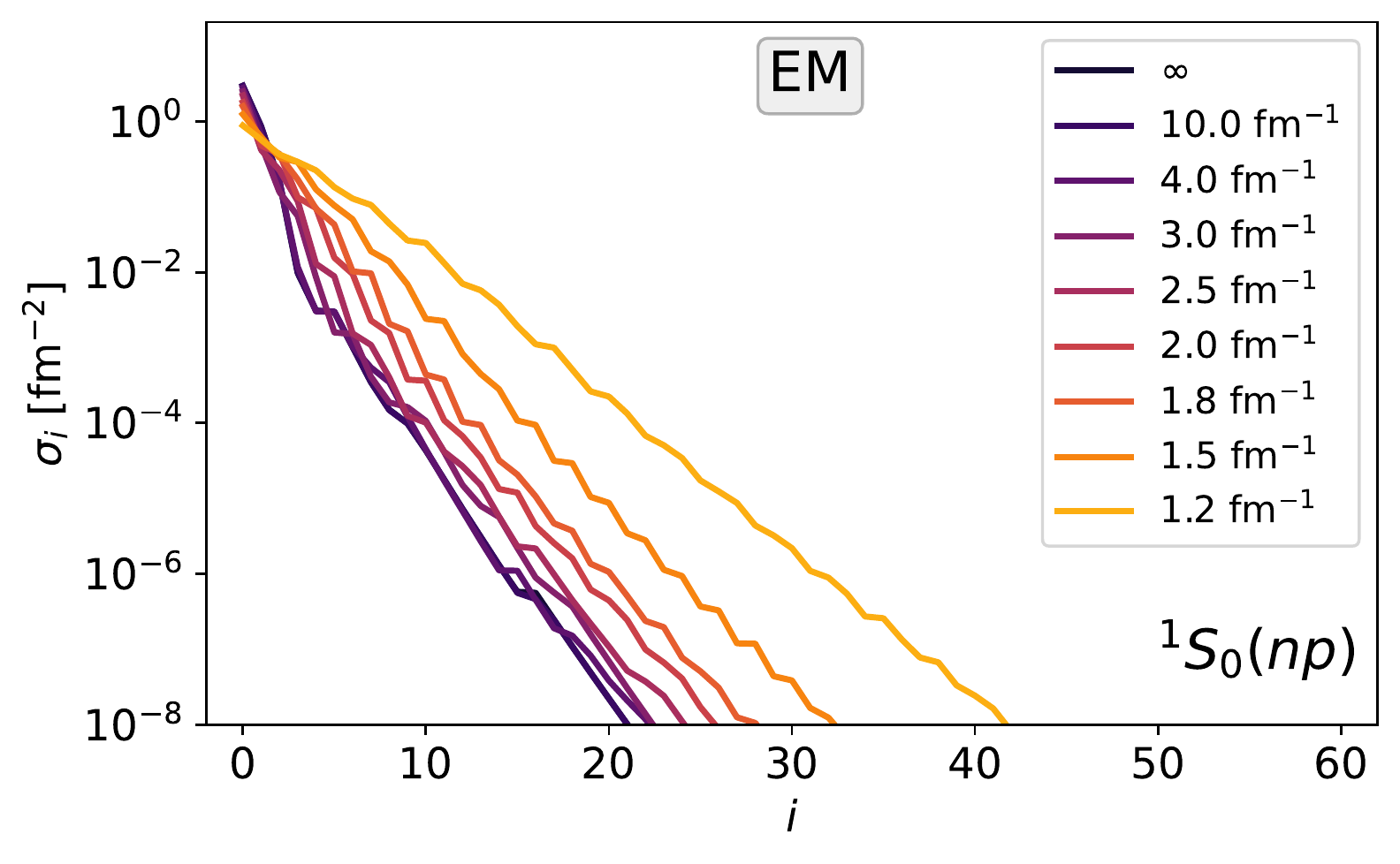}}
  \put(-0.0200,0.0000){\includegraphics[width=\unitlength]{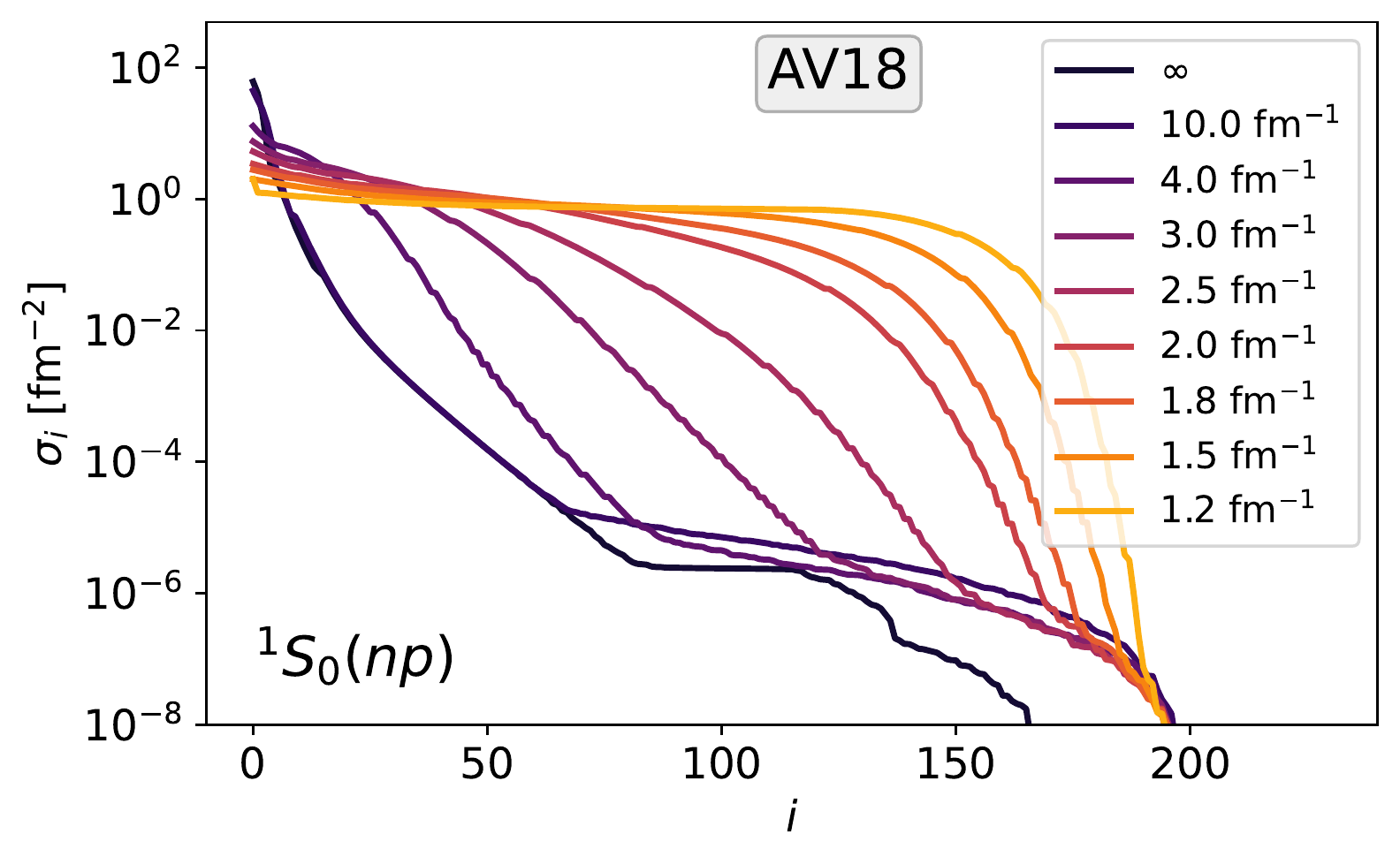}}
\end{picture}
\vspace{-20pt}
\caption{Singular values of the neutron-proton ${}^1S_{0}$ partial wave for the SRG-evolved EM (panel a) and AV18 interactions (panel b) at different resolution scales $\lambda$.}
\label{fig:svd_resolution}
\end{figure}

Next, we consider the effects of a free-space SRG evolution on the singular value spectrum. Figure \ref{fig:svd_resolution}a shows the singular values in the neutron-proton ${}^1S_0$  partial wave for the EM and AV18 interactions, evolved to different resolution scales $\lambda$. Focusing on the former, we note that the number of singular values above the threshold $\sigma_t$ stays changes only weakly as we evolve from the initial interaction to $\lambda=1.8\,\fmi$. Evolving even further to $\lambda=1.5\,\fmi$ and $\lambda=1.2\,\fmi$, the growth of the rank accelerates, and $r$ reaches twice the value of the original interaction. 

At these low resolution scales, momentum transfers (cf.~Eq.~\eqref{eq:srg_momentum_transfer}) that are associated with the dynamics of one-pion exchange start to become suppressed by the decoupling. In a projective RG scheme , we would say that we start to ``integrate out'' the pion \cite{Bogner:2003os,Bogner:2010pq}, but since the SRG is unitary, it cannot destroy interaction strength but only redistribute it. In the two-body sector, the SRG makes $V(s)$ increasingly band-diagonal by sweeping interaction strength towards the diagonal. Eventually, the width of the band becomes small enough that strength starts to push along the diagonal up to higher momenta \cite{Bogner:2010pq}, rendering previously unimportant components of the flowing interaction relevant. 

The growth of the rank is important to keep in mind when we implement the SRG evolution of the SVD factors as described in Sec.~\ref{sec:srgsvd}. We are typically evolving the interaction to resolution scales $\lambda\approx 2.0\fmi$, which have proven to be a sweet spot for nuclear many-body calculations, so we should be able to avoid a dramatic increase in rank. Nevertheless, it seems prudent to ``over-sample'' and include a few extra components in the procedure, so that we can capture the RG flow and do not suffer a loss of unitarity in the two-body system.

The unevolved AV18 interaction, shown in Fig.~\ref{fig:svd_resolution}b, starts out at a much greater rank than the EM interaction due to its much greater extension in momentum space: Due to the hard core of the interaction, AV18 can readily couple incoming and outgoing momenta that differ by as much $20\,\fmi$ \cite{Bogner:2010pq}. The effect of an SRG evolution on the interaction is dramatic: While the most dominant singular values are reduced in size, the rank rapidly expands and the spectrum becomes so flat that possible truncation points for the SVD are between 150 and 200 components. 

\begin{figure}
  \includegraphics[width=\columnwidth]{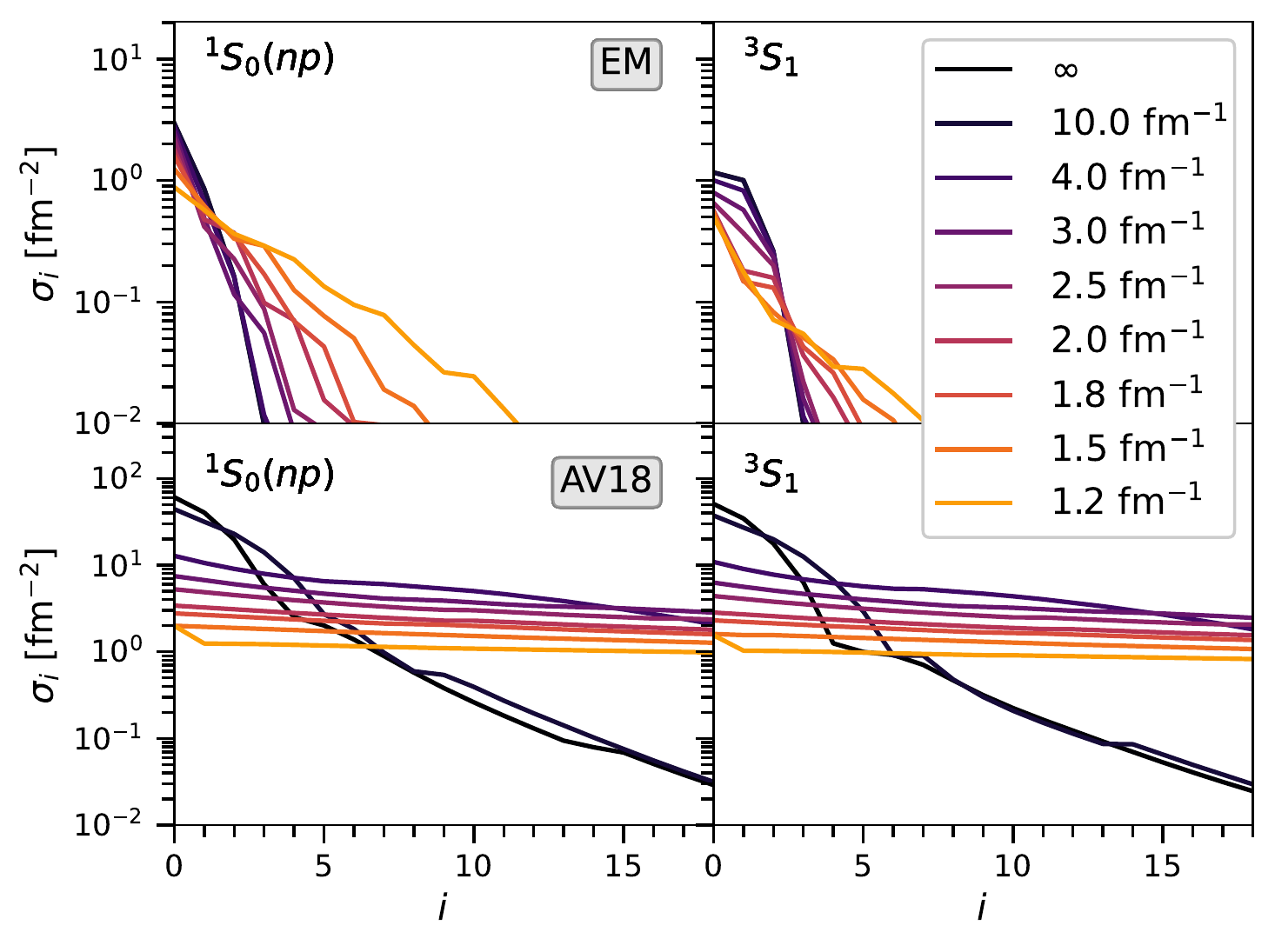}
  \vspace{-20pt}
  \caption{Detail view of the dominant singular values in selected $S$ waves of the EM and AV18 interactions at different SRG resolution scales $\lambda$.}
  \label{fig:svd_detail}
\end{figure}

In Fig.~\ref{fig:svd_detail}, we show a detailed view of the dominant singular values in the neutron-proton $S$ waves of our interactions. For the EM interaction, in particular, we notice that there are just 2--3 particularly dominant singular values before the spectrum drops off rapidly. As we evolve to lower $\lambda$, the relative dominance of just one of these values is enhanced compared to the others, before the growth of the rank eventually becomes a concern beyond $\lambda=1.8\fmi$. This observation is in line with the analysis of Bogner et al. \cite{Bogner:2006qf}, who found that low-rank separable approximations to the $NN$ interaction become \emph{more} accurate as the resolution of an interaction is lowered, since the SVD can be viewed as a generalization of such techniques. 

Curiously, however, the analysis of Ref.~\cite{Bogner:2006qf} reached this conclusion also for a low-resolution $\Vlowk$ interaction that was constructed from AV18 by means of a projective RG decimation \cite{Bogner:2003os,Bogner:2010pq}. While just a few singular values dominate the spectrum at the beginning of our SRG evolution, the rapid flattening of the spectrum and simultaneous growth of the rank appears to be at odds with the $\Vlowk$ result. 

To resolve this issue, we first point out that the size of the singular value is, in general, \emph{not} a sufficient criterion for deciding whether a component of the interaction is relevant for the physics we want to describe or not. Consider a local interaction, whose momentum space matrix elements we can write schematically as
\begin{equation}
  V_\text{reg}(q,q') = V(q,q') f\left(\frac{q-q'}{\lambda}\right) \,,
\end{equation}
where $f$ is a local regulator. In momentum space, the interaction matrix will be band diagonal, and the suggestively named $\lambda$ controls the width of the band. However,$f$ and $\lambda$ do not limit $(q+q')$. Indeed, AV18 is a local potential with a strong repulsive core, and it has large positive matrix elements at high momenta $q$ (see, e.g., Fig.~2 in Ref.~\cite{Bogner:2010pq}). Its \emph{eigenvalues} that are dominated by the high-momentum region exceed the magnitude of the negative eigenvalues from the attractive region at low mometum. The two types of eigenvalues get mixed in the singular value spectrum, which only reflects their absolute value \footnote{Since we are usually working with real symmetric matrices, we can identify what type of eigenvalue a $\sigma_i$ corresponds to by checking whether $\ket{u_i}$ and $\ket{v}_i$ differ by a phase or not.}. 

Additional information that can help us decide whether a component of the interaction is relevant or not resides in the structure of the associated singular vectors. Since we are primarily interested in their behavior under a component-wise SRG evolution, it is useful to consider the expansion of one set of $\ket{u_i(\lambda)}$ (or $\ket{v_i(\lambda)}$) in terms of the set at a different $\lambda$. The expansion coefficients are the entries of the unitary evolution matrix \eqref{eq:unitarySVD} without the restriction that one of the scales is $\lambda=\infty$ (or, equivalently, $s = 0 \,\fm^4$). 

\begin{figure}
  \includegraphics[width=1.02\columnwidth]{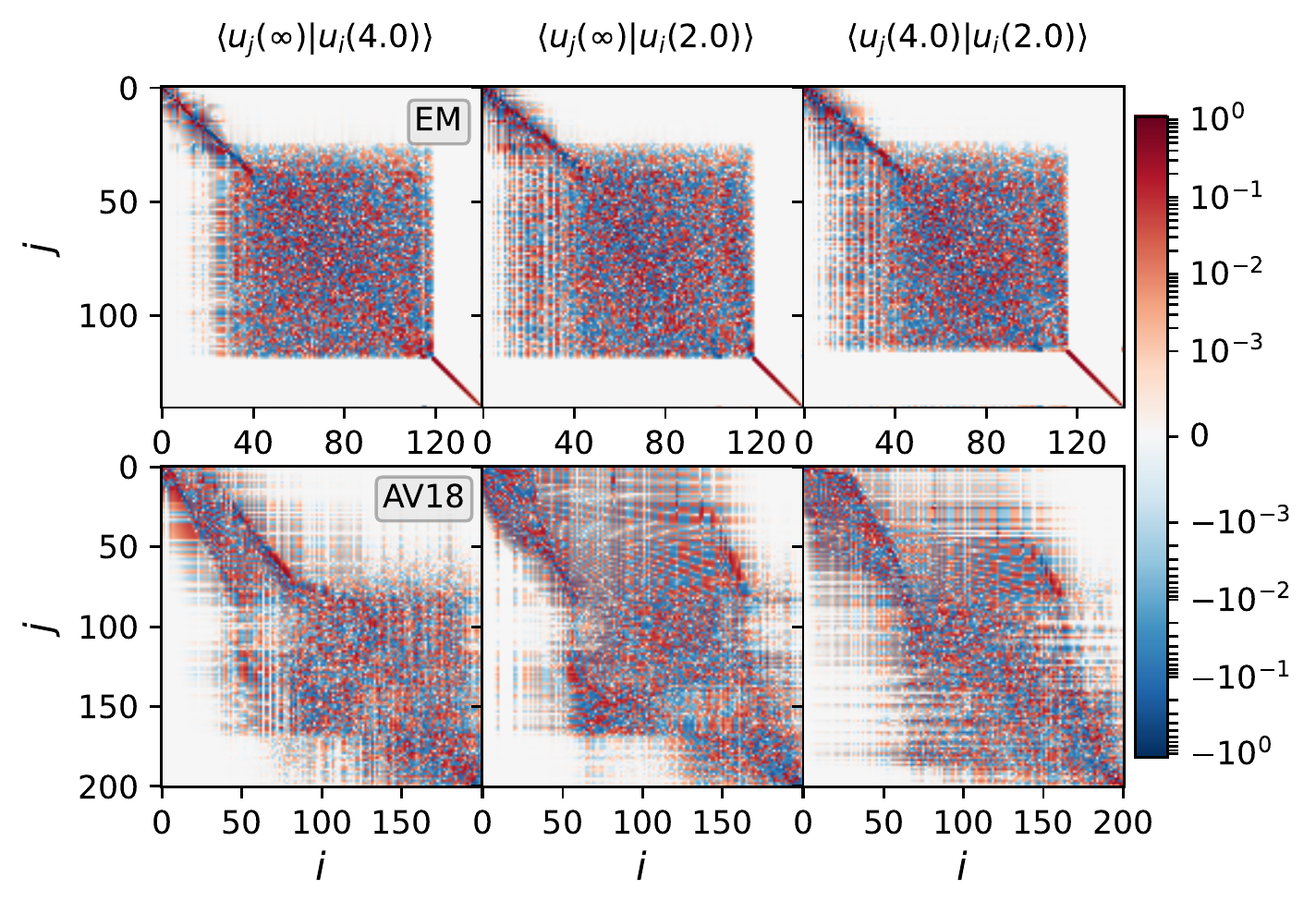}
  \caption{Matrix representation of the unitary transformation $U(\lambda_1,\lambda_2)=\braket{u_j(\lambda_1)}{u_i(\lambda_2)}$ between the initial and final resolution scales $\lambda_1,\lambda_2 \in \{2\,\fmi, 4\,\fmi, \infty\}$ (cf.~\eqref{eq:unitarySVD}). Here, we specifically focus on the ${}^1S_0(np)$ partial wave. The basis vectors are ordered by decreasing singular value. Note the logarithmic color scale.}
  \label{fig:svd_U}
\end{figure}

In Fig.~\ref{fig:svd_U}, we show these matrices for $(\lambda_1,\lambda_2) = (\infty, 4.0\,\fmi), (\infty, 2.0\,\fmi)$ and $(4.0\,\fmi, 2.0\,\fmi)$. Considering the transformation for the EM interaction first, we notice that the matrices all look very similar: There is a somewhat washed out diagonal band in the upper left corner, a large central block, and a very sharp diagonal in the lower right corner. It is rather straightforward to relate them to our observations for the interaction's truncated SVD. Based on Fig.~\ref{fig:svd_resolution}, the upper left and central blocks contain the $\ket{u_i}$ of the singular values $\sigma_i \gtrsim 10^{-6}$ and $\sigma_i\lesssim 10^{-6}$, respectively. The particular value at which the separation between the blocks occurs is most likely controlled by the accuracy settings of the ODE solver we use. The lower right block contains the singular vectors that do not evolve, which implies that we have $V = 0$ and $H = T$. The block structure is a consequence of the nonlocal regularization of the interaction, which suppresses the momentum space matrix elements independently in the incoming and outgoing momenta as
\begin{equation}
    V_\text{reg}(q,q') = V(q,q')e^{-\left(\frac{q}{\Lambda}\right)^{2n}}e^{-\left(\frac{q'}{\Lambda}\right)^{2n}}
\end{equation}
where $n=2$ or $n=3$ and $\Lambda=2.5\,\fmi$ \cite{Entem:2003th}. Because of the structure of the matrix, the first $r$ singular vectors remain almost completely decoupled from the rest of the spectrum when we evolve, although we note that the central block stretches out to the lower indices in the $U(\infty,2.0)$ matrix, which is in line with the slight growth of the rank. This effect is slightly less pronounced in $U(4.0,2.0)$ because the two evolved bases are more similar to each other.

For AV18, the structure of the unitary transformations is much more complex. The components that will eventually be most relevant for the low-momentum sector are scattered throughout the basis and difficult to identify a priori. As we evolve, the matrices $U(\infty,\lambda)$ actually become \emph{less} structured, which reflects the growth of the rank in Fig.~\ref{fig:svd_resolution}. The matrices suggest that we can anticipate that between 160 and 170 out of 200 components of the interaction are necessary to implement the SRG using the factorized form of the interaction. There is, however, a simplification in the structure of the matrix $U(4.0,2.0)$ --- i.e., if we perform the SVD at that $\lambda=4.0\,\fmi$, we may be able to observe an improvement of the low-rank structure because it would be safe to project out the high momentum components that are already decoupled, as in \Vlowk. We will see evidence of this in the next section, when we discuss the impact of this factorized evolution on the AV18 deuteron ground-state energy (cf.~Fig.~\ref{fig:deuteron_av18}).  

The takeaway message from our investigation is that in general, the size of the singular values is only a necessary but not a sufficient criterion for deciding which singular vectors are required for an accurate low-rank representation of the interaction in the low-momentum sector. Additional criteria that reflect the momentum structure of the singular vectors may have to be taken into account. For chiral interactions with nonlocal regulators, the selection based on the singular values works because we do not have many strongly positive eigenvalues if the initial cutoff is not too high. For locally regularized interactions, a reduction to low rank only works if there is a limit on $(q+q')$ as well. Clearly, this is not the case for AV18. 

\begin{figure*}
\includegraphics[width=0.98\textwidth]{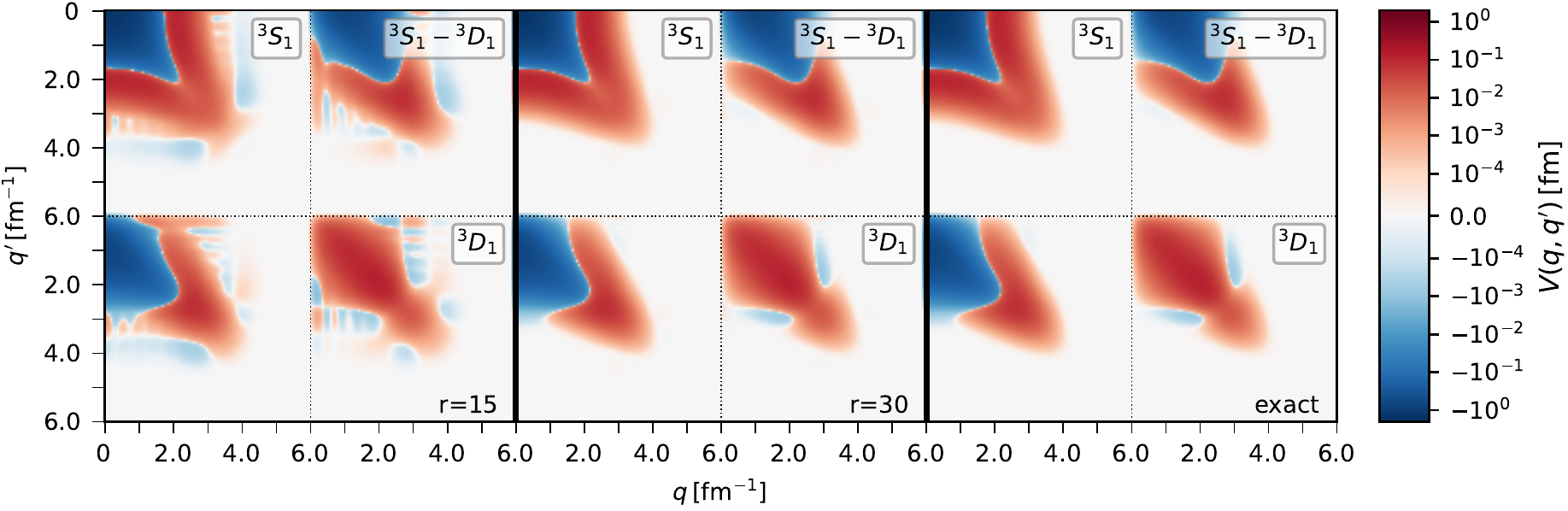}
\caption{Momentum-space matrix elements of the EM interaction at $\lambda = 2.0\,\fmi$ in the deuteron channel.  The matrix elements shown in the first two panels were generated using $15$ and $30$ components per partial wave, respectively. The panel on the right shows the full matrix-based evolution. The matrix elements are given in scattering units ($\hbar = c = \hbar^2/ m = 1$). Also note the logarithmic color scale.  }
\label{fig:potential_deuteron}
\end{figure*}

\begin{figure*}
\centering
\includegraphics[width=0.98\textwidth]{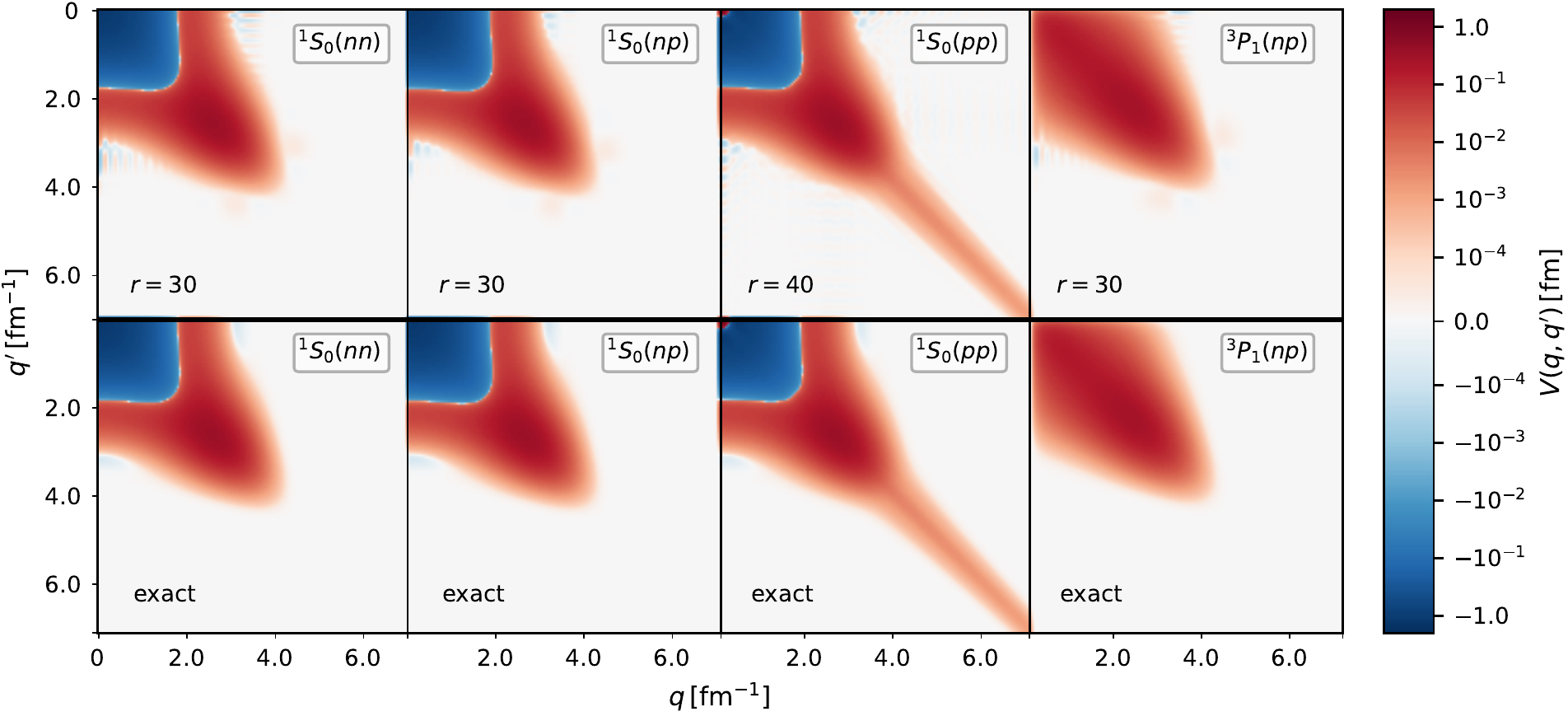}
\caption{Momentum-space matrix elements of the EM interaction at $\lambda = 2.0\text{ fm}^{-1}$ in selected uncoupled partial waves.  The top panels are by SVD-SRG, the bottom panels by conventional matrix-based SRG evolution. The matrix elements are given in scattering units ($\hbar = c = \hbar^2 / m = 1$). Note the logarithmic color scale.}
\label{fig:potential_single}
\end{figure*}

\subsection{SVD-Based SRG Evolution}
\label{sec:test_svdsrg}
Let us now implement the SRG evolution of an SVD-factorized interaction --- SVD-SRG, for short --- using the formalism developed in Sec.~\ref{sec:srgsvd}. As a first example, we evolve the SVD of the EM interaction to $\lambda=2.0\,\fmi$. In Fig.~\ref{fig:potential_deuteron}, we show the momentum space matrix elements in the deuteron channel for different SVD ranks. Using only $15$ components per partial wave, we still observe some distortion, but for $30$ components, the evolution agrees with the result from the evolution of the full matrix within absolute deviations on the order of $10^{-4}$ that one can identify upon scrutiny. It is worth noting that we did not attempt to fine-tune $r$ using the information from previous sections (e.g., Fig.~\ref{fig:num_of_comp}) --- the decision to use the same rank for all partial waves was purely for convenience. 

In Fig.~ \ref{fig:potential_single}, we investigate the performance of the SVD-SRG in other selected partial waves. The matrix elements for the neutron-neutron and neutron-partial waves were obtained using 30 components, while $40$ components had to be used for the ${}^1S_0$ proton-proton partial wave due to the presence of the Coulomb interaction --- note the Coulomb tail along the diagonal, which is absent in the other isospin channels. We can again note some very weak oscillations and ``fraying'' around the edges of the main structures, but the absolute values of these deviations are again on the order of $10^{-4}$ or below. 

\begin{figure}
  \setlength{\unitlength}{0.5\columnwidth}
  \begin{picture}(2.0000,4.0000)
    \put(0.0000,3.0000){\includegraphics[height=0.98\unitlength]{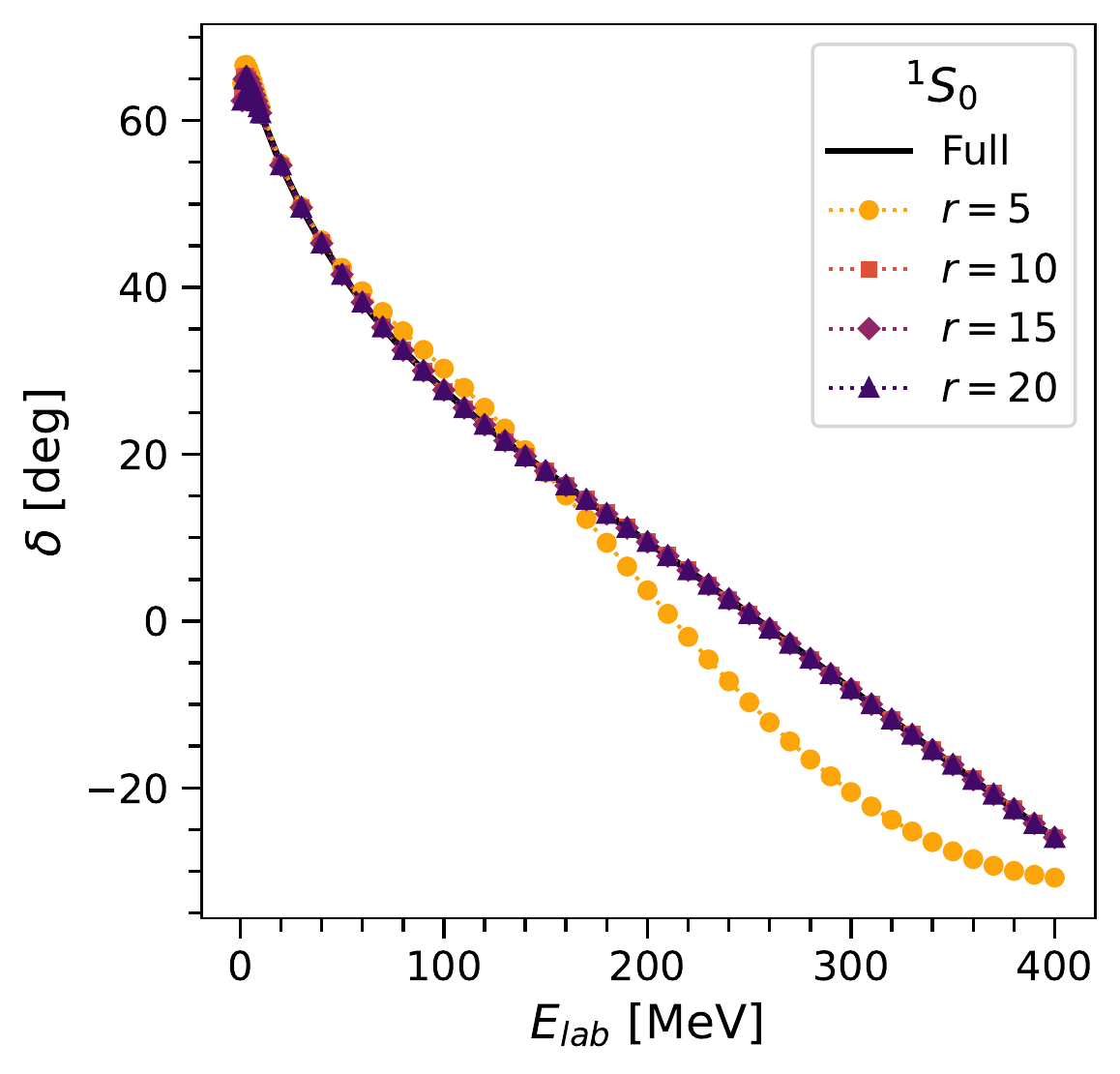}}
    \put(1.0000,3.0000){\includegraphics[height=0.98\unitlength]{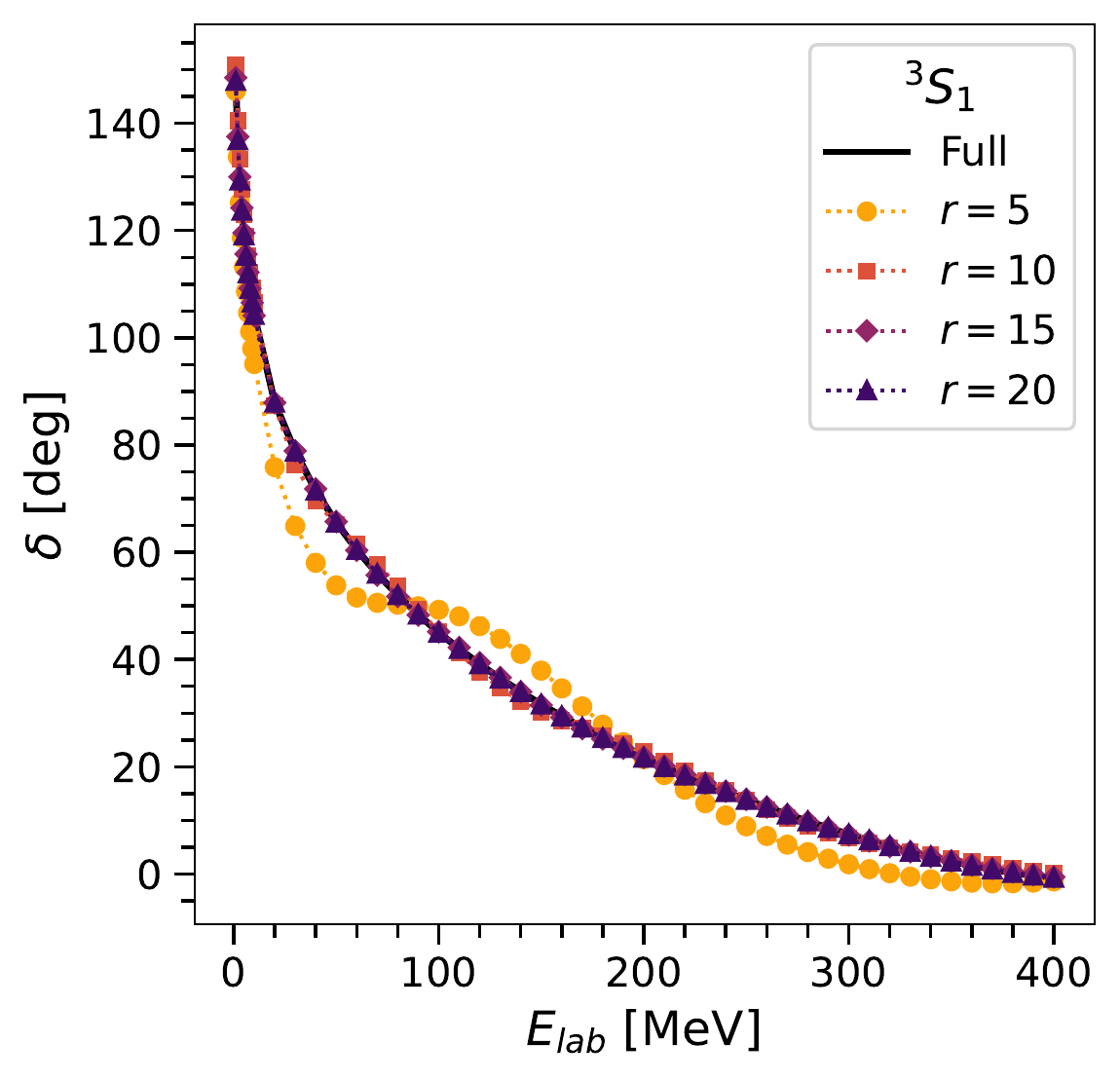}}
    \put(0.0000,2.0000){\includegraphics[height=0.98\unitlength]{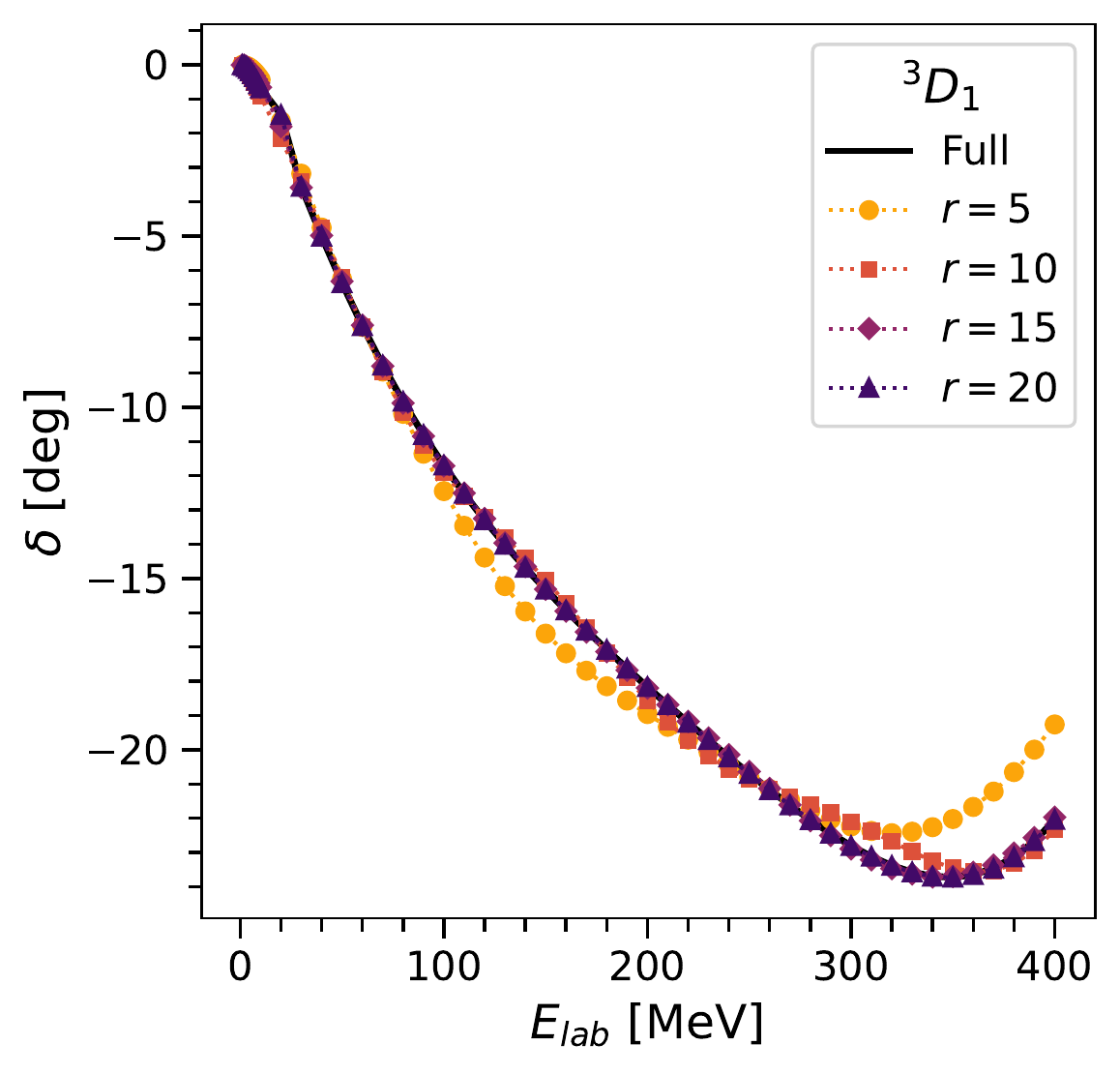}}
    \put(1.0000,2.0000){\includegraphics[height=0.98\unitlength]{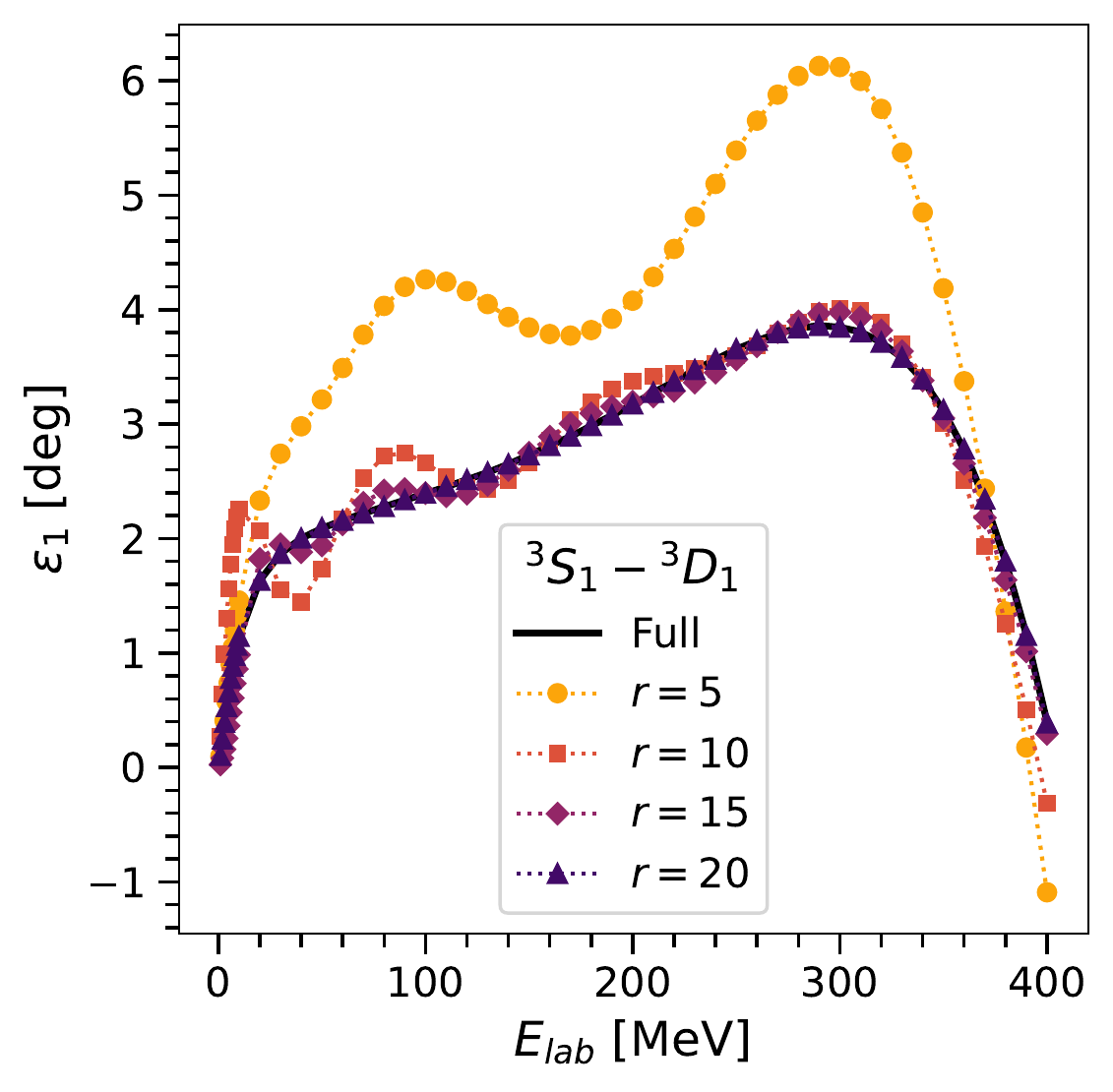}}
    \put(0.0000,1.0000){\includegraphics[height=0.98\unitlength]{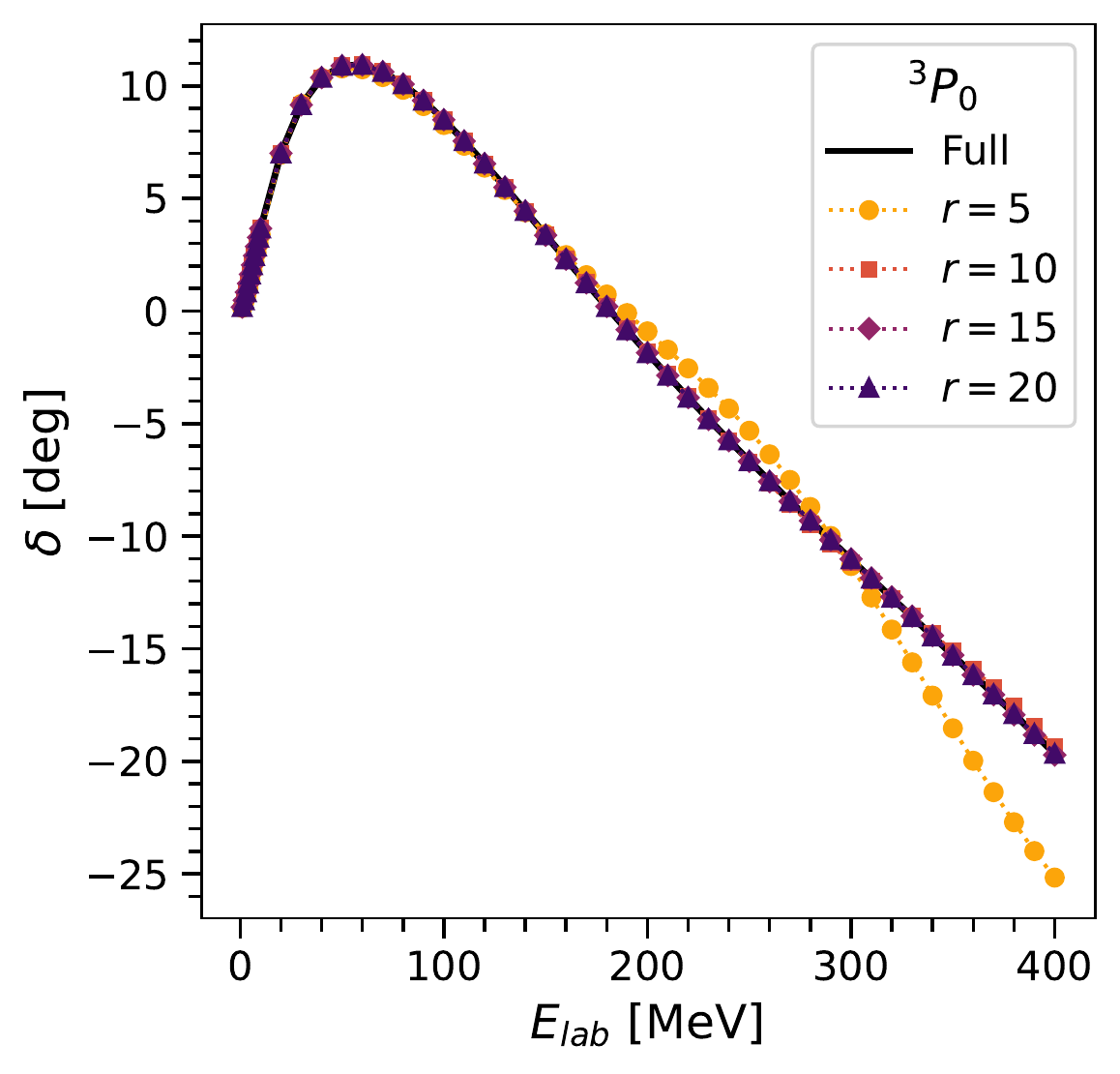}}
    \put(1.0000,1.0000){\includegraphics[height=0.98\unitlength]{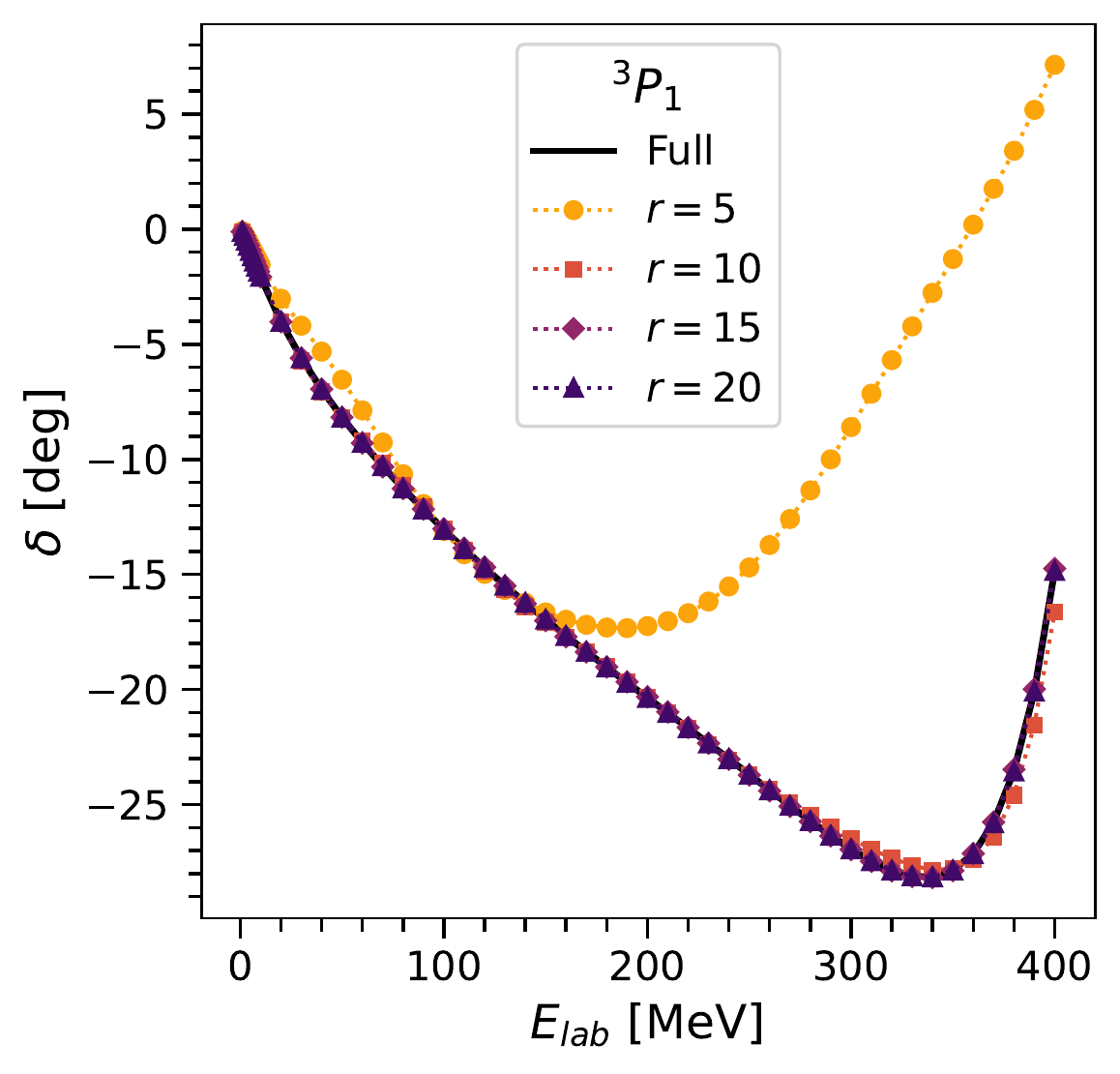}}
    \put(0.0000,0.0000){\includegraphics[height=0.98\unitlength]{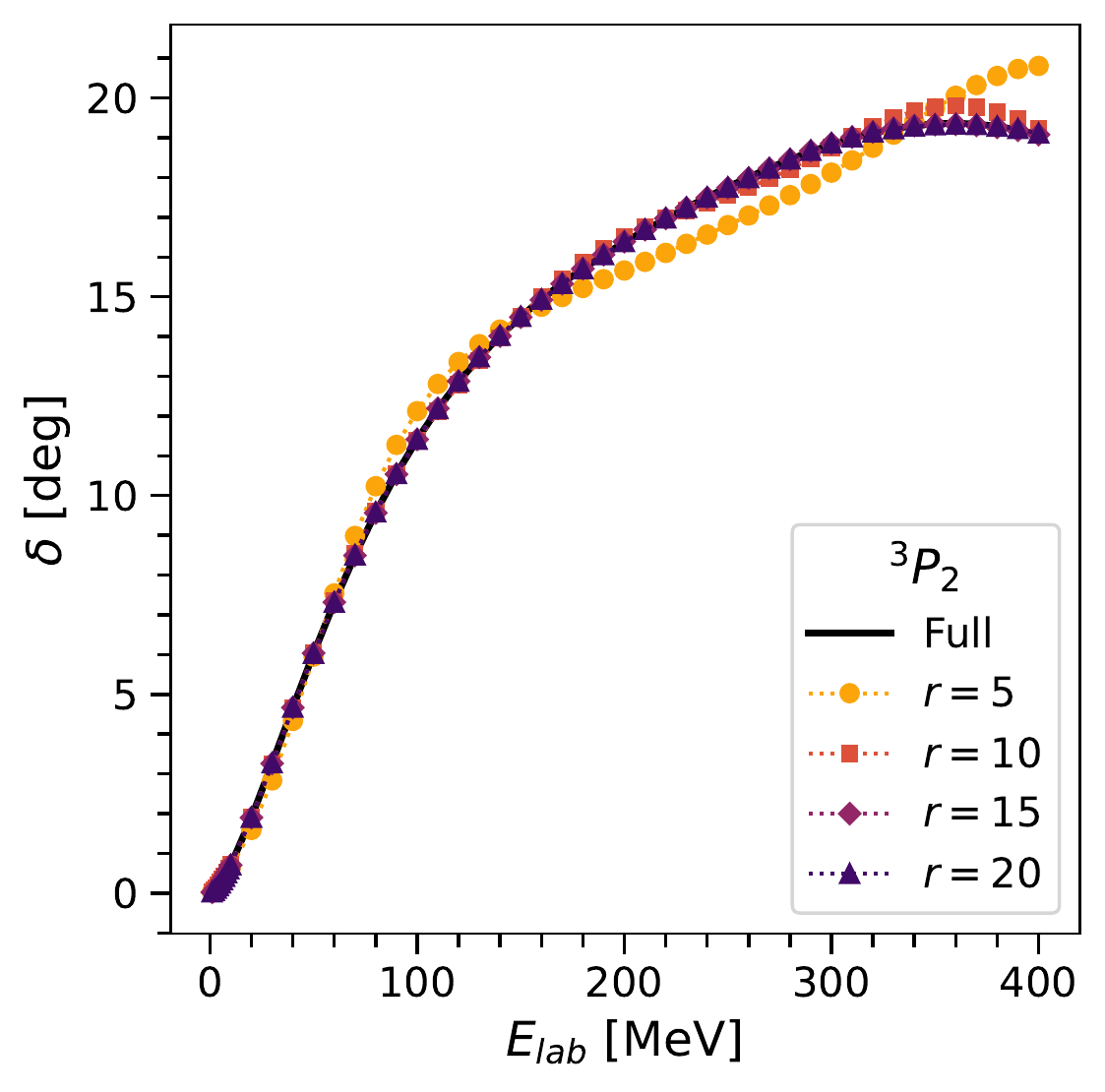}}
    \put(1.0000,0.0000){\includegraphics[height=0.98\unitlength]{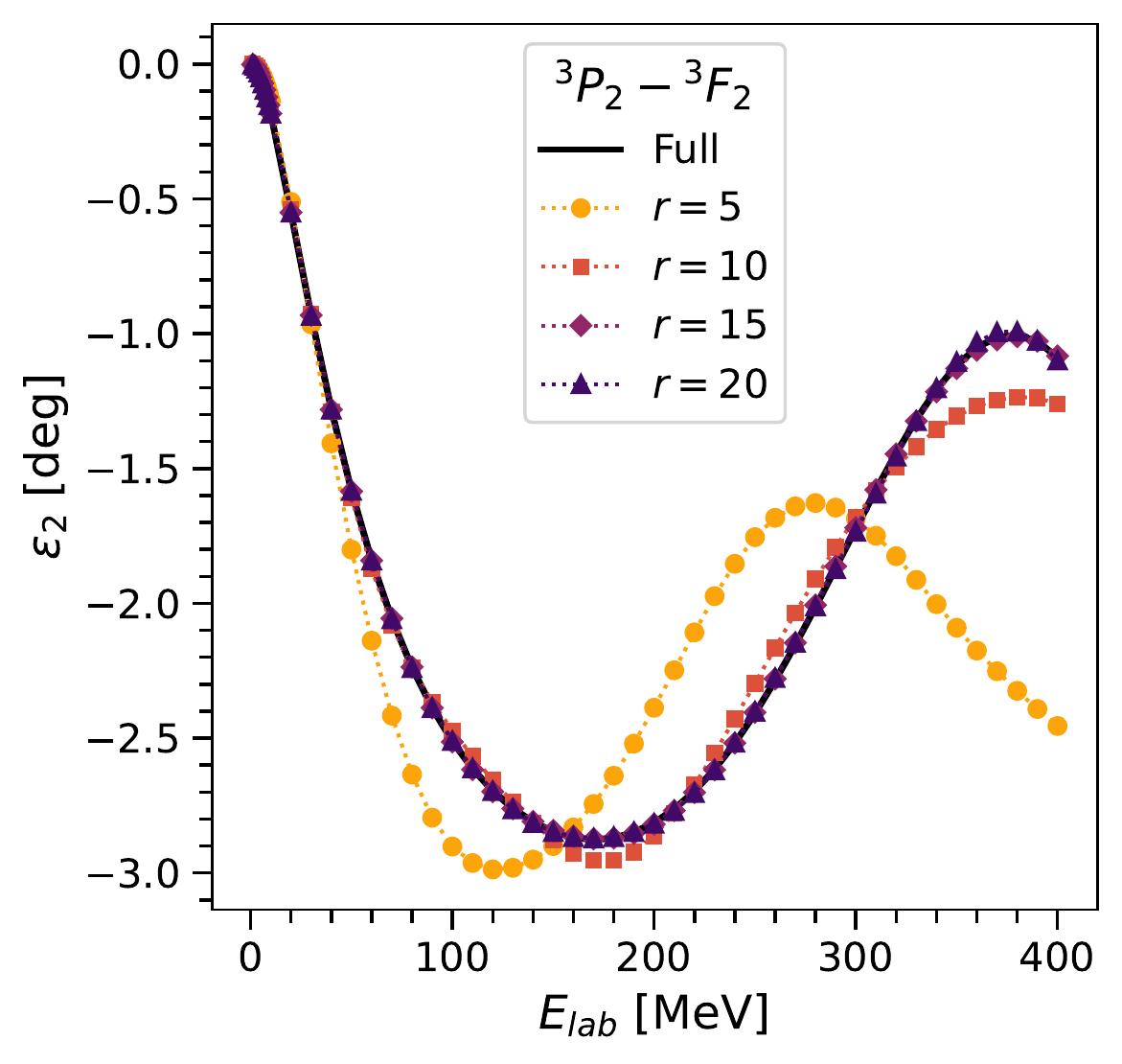}}
  \end{picture}
  \caption{Selected neutron-proton phase shifts and mixing angles of the EM interaction resolution $\lambda=2.0\fmi$. The SVD-SRG evolution for different ranks is compared to the matrix-based evolution, which exactly preserves the phase shifts of the unevolved EM interaction by construction.}
  \label{fig:np_phaseshift}
\end{figure}

Since the SVD-SRG seems to work accurately on the matrix element level, we now use the factorized interactions to compute observables in the nucleon-nucleon system, namely scattering phase shifts and the deuteron binding energy. In Fig.~\ref{fig:np_phaseshift}, we show the neutron-proton phase shifts and mixing parameter of the SVD-SRG evolved EM interaction in the deuteron channel as well as other selected partial waves. As we can see, between 5 and 10 components of the interaction are actually sufficient to reach agreement with the conventional matrix-based evolution, only the ${}^3S_1-{}^3D_1$ mixing angle $\varepsilon_1$ seems to require a few additional components. These results match our expectations based on the SVD of the initial interaction (cf.~Fig.~\ref{fig:num_of_comp}) and the need to accommodate a slight growth in the rank as we evolve, here to $\lambda=2.0\fmi$.

\begin{figure}
  \setlength{\unitlength}{0.5\columnwidth}
  \begin{picture}(2.0000,2.0000)
    \put(0.0000,1.0000){\includegraphics[height=0.98\unitlength]{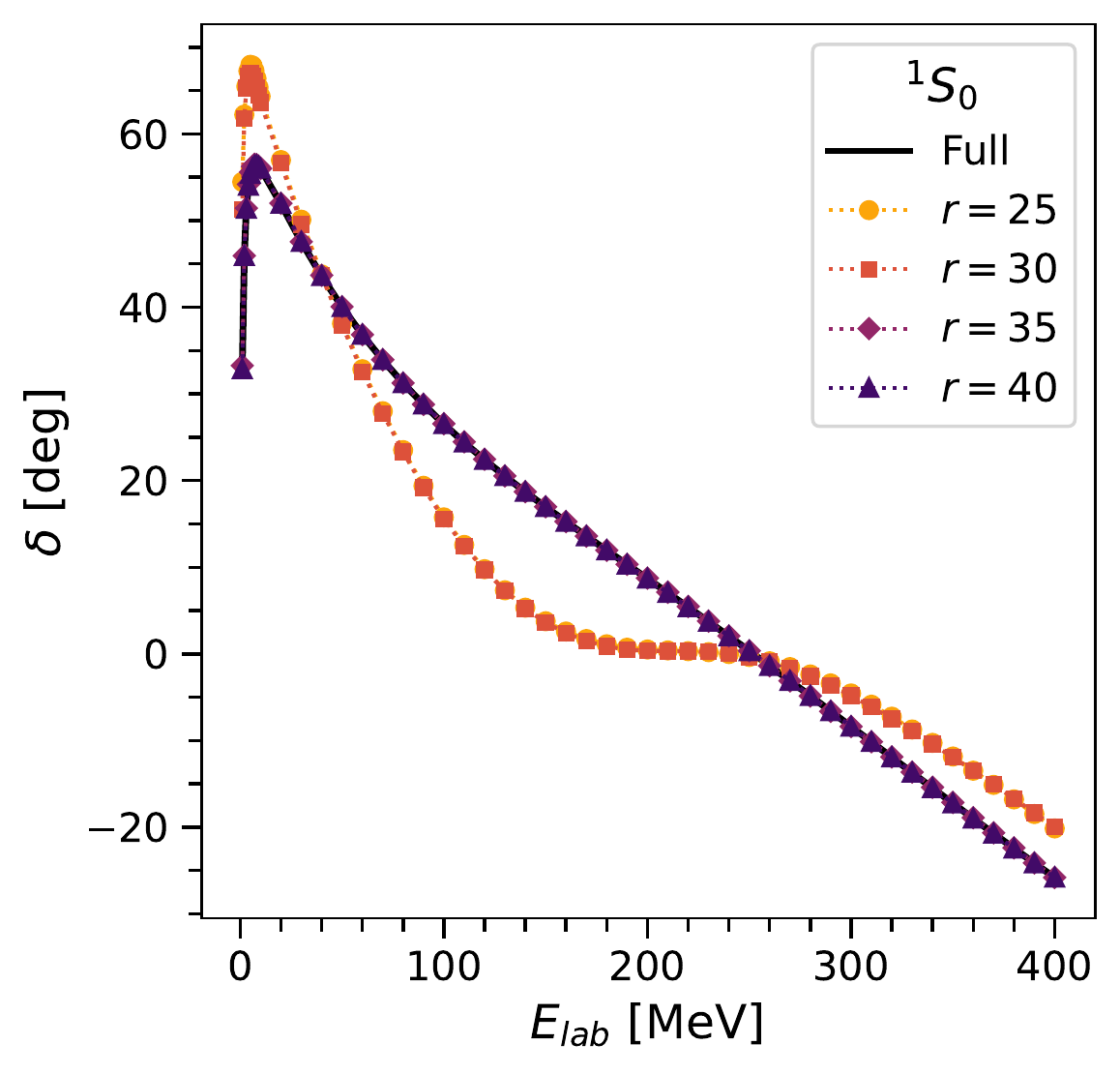}}
    \put(1.0000,1.0000){\includegraphics[height=0.98\unitlength]{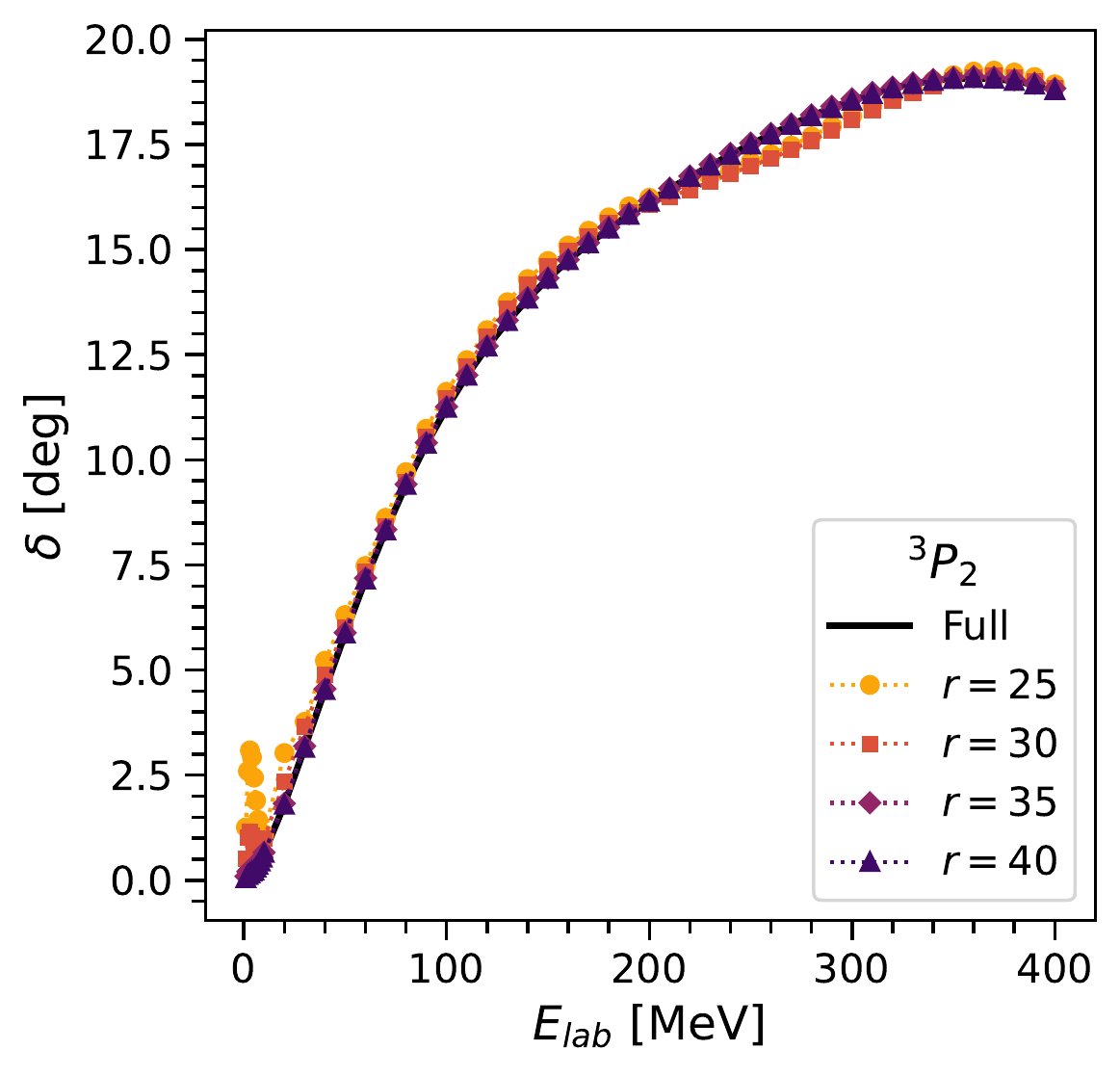}}
    \put(0.0000,0.0000){\includegraphics[height=0.98\unitlength]{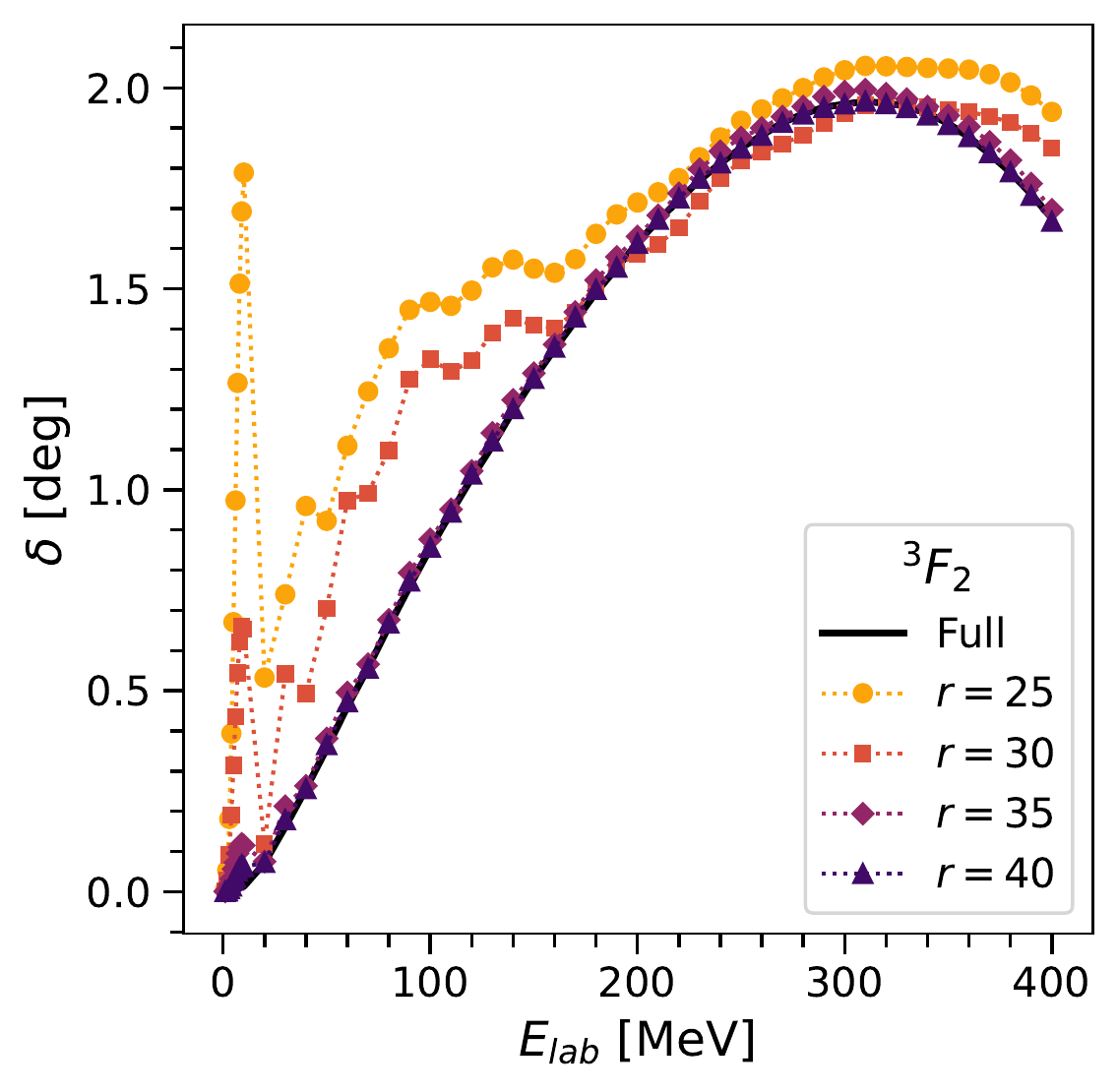}}
    \put(1.0000,0.0000){\includegraphics[height=0.98\unitlength]{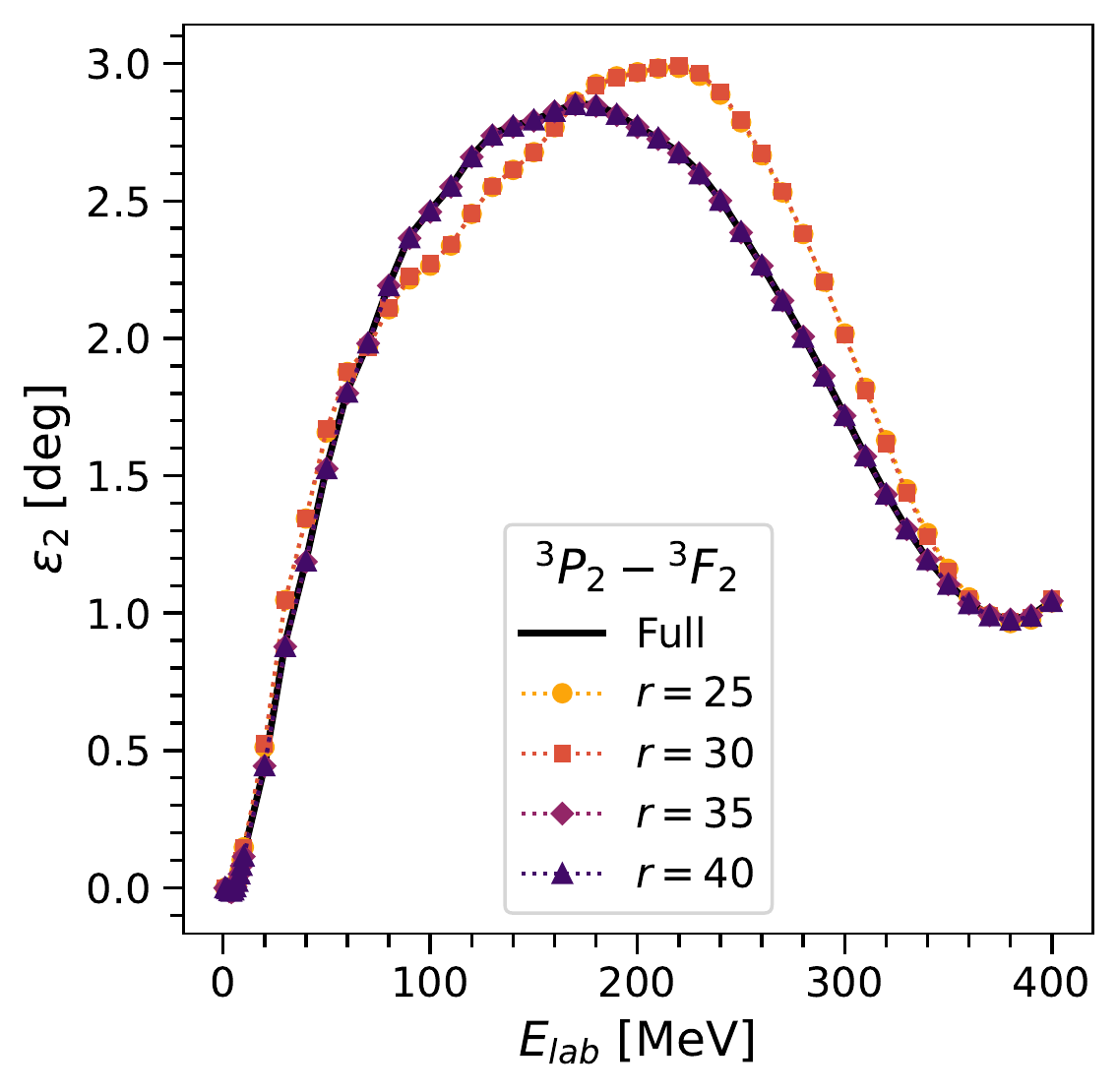}}
  \end{picture}
  \caption{Selected proton-proton phase shifts and mixing angles of the EM interaction resolution $\lambda=2.0\fmi$.
  Note the higher ranks for the SVD-SRG evolution compared to Fig.~\ref{fig:np_phaseshift}.}
  \label{fig:pp_phaseshift}
\end{figure}

In the proton-proton channel, the proper treatment of the Coulomb interaction forces us to increase the number of components to $30-35$, as we can see from the partial wave and mixing angles in Fig.~\ref{fig:pp_phaseshift}. This is particularly evident since the deviations for rank $r<30$ primarily occur at low energy (long distances), which are governed by $V_C$. This includes the oscillation of the ${}^1S_0$ phase shift between $E_\text{lab}=100--250\,\MeV$, which is tied to the exaggerated $\delta$ in the low-energy region.

\begin{figure}
  \includegraphics[width=0.48\textwidth]{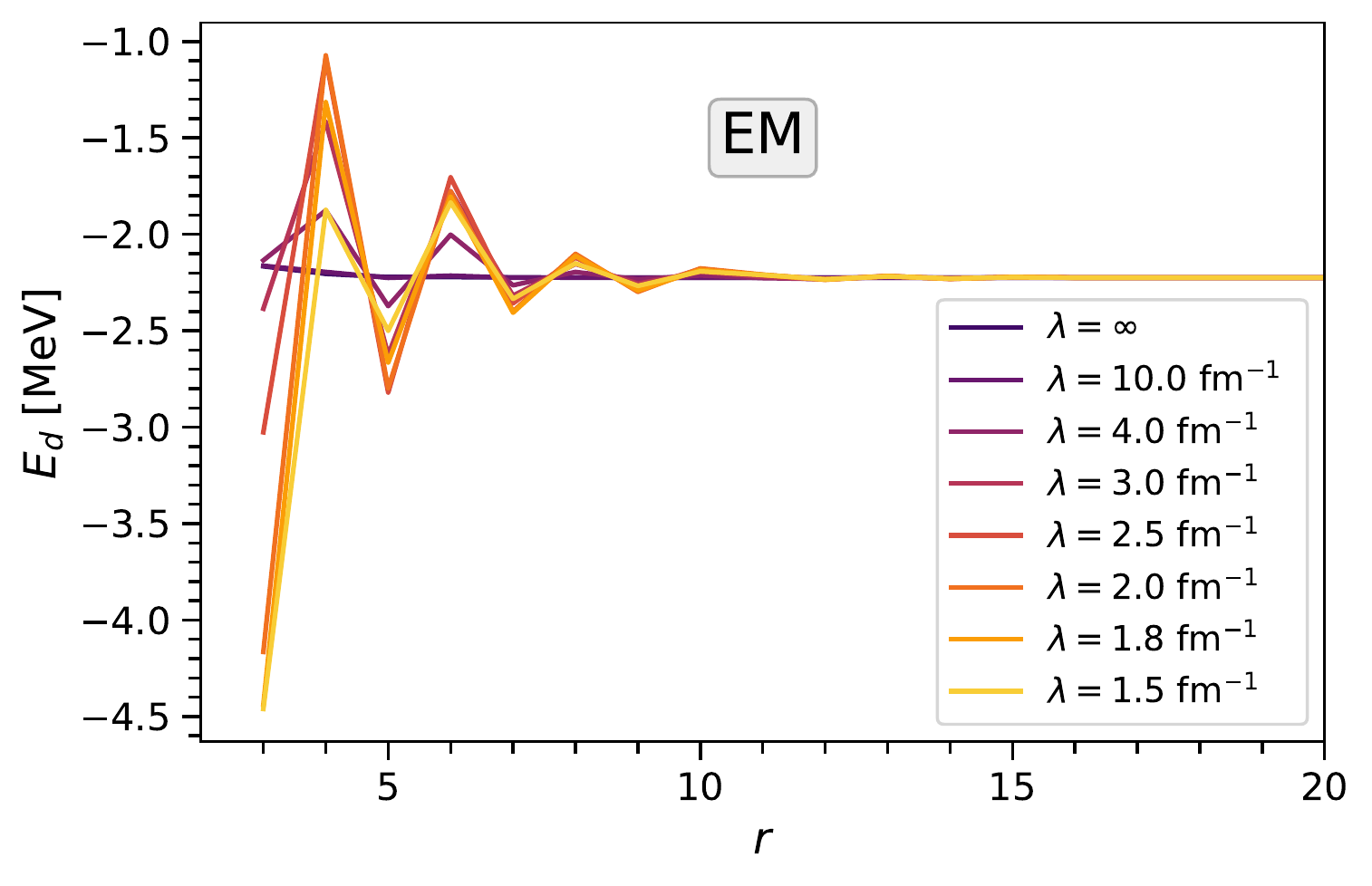}
  \vspace{-20pt}
  \caption{Ground-state energy of the deuteron for the SVD-SRG-evolved EM interaction at different resolution scales $\lambda$. Here, $r$ refers to the number of components per partial wave.}
  \label{fig:deuteron_chi2b}
\end{figure}

Next, we study the SVD-SRG evolution of the deuteron ground-state energy $E_d$, which is shown as a function of the rank (per partial wave) and the resolution scale in Fig.~\ref{fig:deuteron_chi2b}. Since $E_d$ must be invariant under unitary transformations in the two-body system, the curves for different $\lambda$ must collapse once we have included a sufficient amount of components. At $r=12$, the differences from the exact result are in the single-$\keV$ range, which is expected based on our observation that the mixing angle $\varepsilon_1$ converges a bit more slowly in $r$ than other neutron-proton scattering quantities discussed above (cf.~Fig.~\ref{fig:np_phaseshift}). At lower ranks $r$, we note that the artifacts that spoil the unitarity of the evolution get worse as $\lambda$ decreases, which is due to the accelerating growth of the interaction's rank as the SRG decouples the long-range pion physics (cf.~Sec.~\ref{sec:sv_spectra}). 

\begin{figure}
  \includegraphics[width=0.48\textwidth]{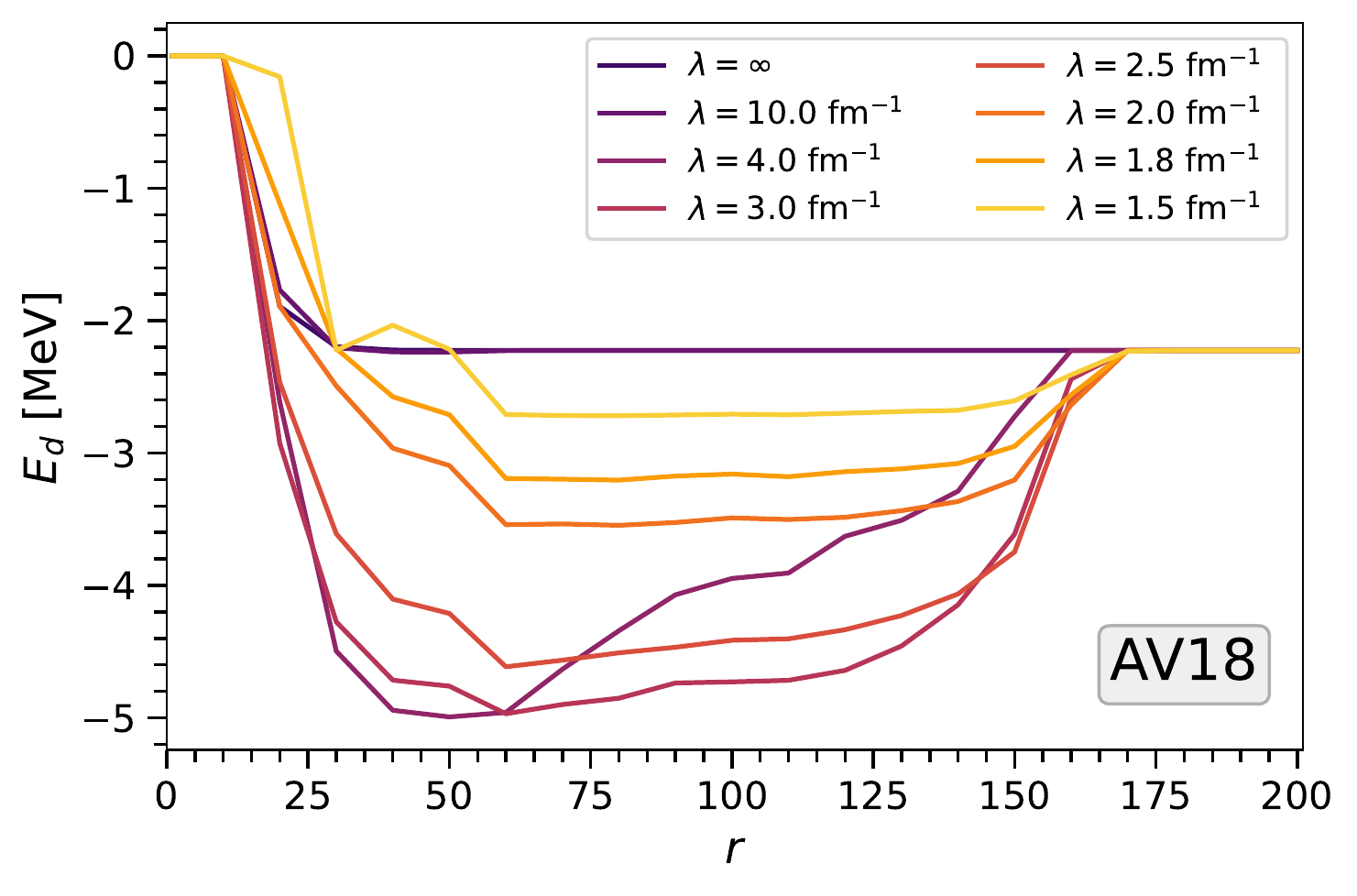}
  \vspace{-20pt}
  \caption{Ground-state energy of the deuteron for the SRG-evolved rank-$r$ approximation of AV18 at different resolution scales $\lambda$.}
  \label{fig:deuteron_av18}
\end{figure}

Overall, the results for the SVD-SRG evolution of the EM interaction show that the method is well-behaved for chiral $NN$ interactions, or at least for those that employ nonlocal regularization schemes. Based on our observations in Sec.~\ref{sec:sv_spectra}, it does not come as a surprise that the story is very different for a hard interaction like AV18. If we truncate the SVD based on the size of AV18's singular values, we need to keep almost all components to reproduce the observables in the two-nucleon system. For illustration, we show the AV18 deuteron ground-state energy in Fig.~\ref{fig:deuteron_av18}: About 170 singular components are necessary to ensure the invariance of $E_d$. At lower resolution $\lambda$, there are plateaus that indicate that many components of the interaction no longer contribute to $E_d$ due to the decoupling of the momentum scales. The fact that these plateaus become more pronounced as $\lambda$ decreases means that low-rank approximations to the evolved AV18 interaction become more accurate, as observed for the projective $\Vlowk$ approach by Bogner et al. \cite{Bogner:2006qf}. Unfortunately, the structure of the unitary transformation $\eqref{eq:unitarySVD}$ (cf. ~Fig.~\ref{fig:svd_U}) is too complex to allow a restriction of the evolution to only these components early on in the flow.

\subsection{Harmonic Oscillator Basis}
\label{sec:ho}

At some stage in the preparation of nuclear interactions for configuration space many-body methods, (spherical) harmonic oscillator (HO) bases comes into play. This is primarily due to the fact that it allows an exact separation of the center-of-mass and intrinsic degrees of freedom in the many-body states if one works in a so-called $\EMax$-complete Hilbert spaces \footnote{Other authors use a different symbols for the total oscillator energy quantum number, e.g., $\mathcal{N}_\text{max}$, and refer to the truncation accordingly. The definition of the truncation is otherwise unchanged.}, where $\EMax=\sum_i (2 n_i + l_i)$ characterizes the total energy of the oscillator state (see, e.g., \cite{Barrett:2013oq,Hergert:2016jk}). In the context of the present work, we can either change the basis of our singular vectors from momentum to HO states via a unitary transformation, or implement the SVD-SRG directly in HO representation. The latter option is of practical interest: The SRG evolution of three-nucleon forces is easier to implement in an $\EMax$-complete Jacobi-HO representation than in momentum representation because the antisymmetrization operator has a block-diagonal structure in the former \cite{Hebeler:2012ly,Hebeler:2020ex,Nogga:2006yg,Roth:2011kx,Roth:2014fk,Jurgenson:2009bs,Jurgenson:2011zr}.

\begin{figure}
    \centering
    \includegraphics[width=0.48\textwidth]{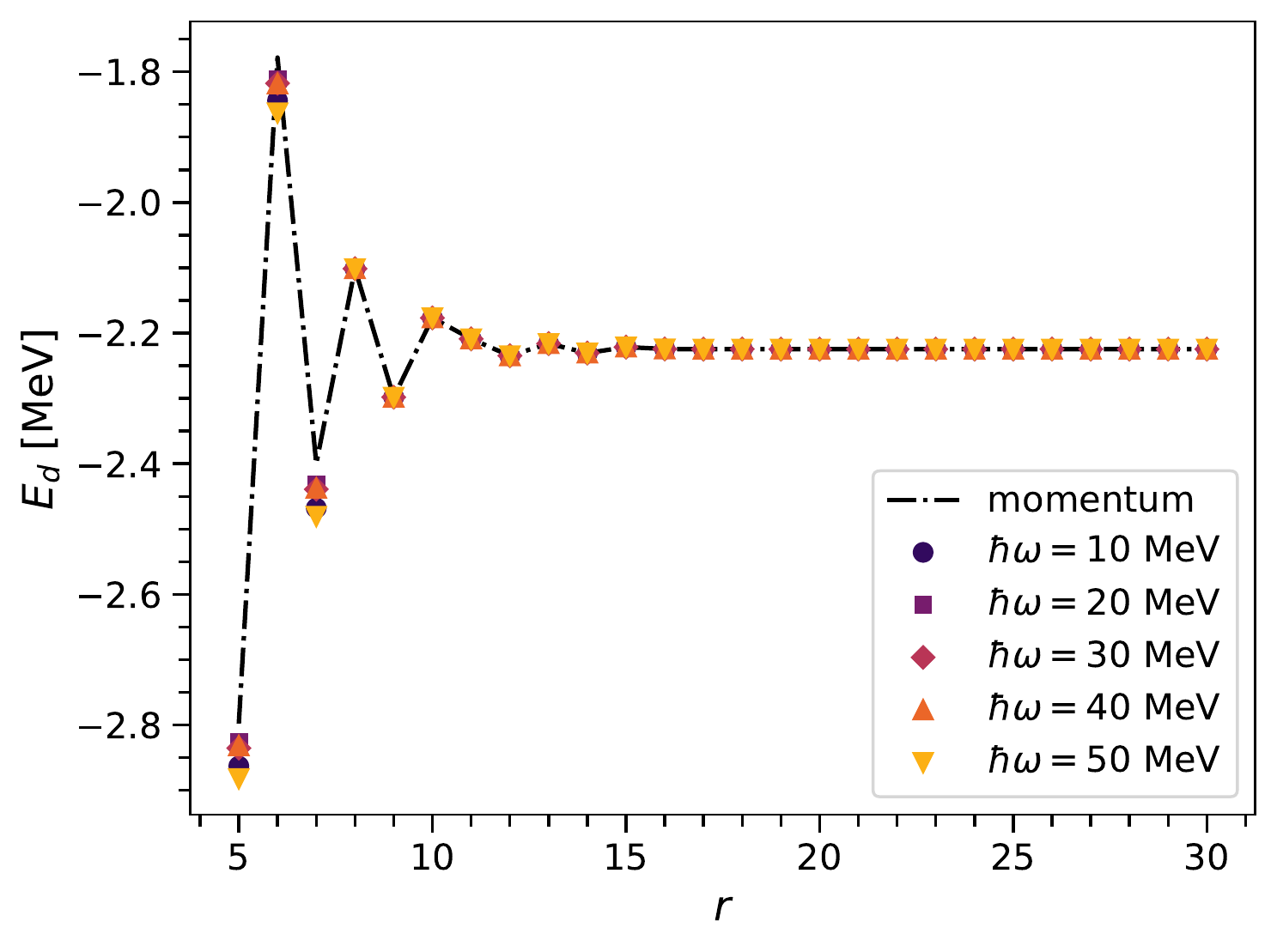}
    \vspace{-20pt}
    \caption{Deuteron ground-state energy of the SVD-SRG evolved EM interaction at $\lambda=2\,\fmi$ as a function of the rank. The SVD-SRG and subsequent diagonalization are performed in relative HO bases with different $\hbar\omega$.}
    \label{fig:deuteron_ho}
\end{figure}

In the present work, we have implemented both approaches and validated that they give consistent results for the deuteron ground-state energy. This is demonstrated for the SVD-SRG evolved EM interaction at $\lambda=2.0\,\fmi$ in Fig.~\ref{fig:deuteron_ho}. Analogous to Fig.~\ref{fig:deuteron_chi2b}, $E_d$ should become invariant under SVD-SRG evolution once a sufficiently high rank is reached. Similar to the momentum-space SVD-SRG, that rank is $r\approx 15$, independent of the HO energy scale $\hbar\omega$. For $r<15$, the behavior of $E_d$ is also the same as for the momentum-space SVD-SRG, which is shown for comparison (also cf.~Fig.~\ref{fig:deuteron_chi2b}). We note that the size of the deviations from the exact value has a weak dependence on $\hbar\omega$, and that the deviations from the momentum space curve are greater for the lowest and largest choices. These choices amount to a tuning of the infrared and ultraviolet ``cutoffs'' of the finite HO basis to the scales of the problem (here, the deuteron wave function). This subject has been explored extensively in the context of large-basis extrapolations in recent years \cite{Coon:2012uq,More:2013bh,Konig:2014fk,Furnstahl:2015sf,Wendt:2015zl,Odell:2016qq}.

For our purposes, the main takeaway message is that the SVD-SRG in HO representation seems to perform as well as the momentum-spaced framework, and that the conclusions regarding the rank of nuclear interactions remain valid, for better (EM) or worse (AV18). The representations of $V$ in the momentum and HO bases are reasonably similar, and while $T$ is tridiagonal rather than diagonal in HO representation, it can still only connect basis states that are energetically close. Consequently, the structure and action of the generator will be very similar as well.

\section{Transformation to the Laboratory Frame}
\label{sec:talmi}

Moving on from exploring the SVD and SVD-SRG in the two-body system, our next goal is to apply the factorized interactions in many-body calculations. To do so, we need to consider the transformation of the interaction from the center-of-mass frame to the laboratory frame. This involves the Talmi-Moshinsky transformation from the intrinsic (i.e., center-of-mass plus Jacobi HO) and laboratory frames (see, e.g., \cite{Kamuntavicius:2001kr,Moshinsky:1959cp}).
The singular vectors are coupled to the center-of-mass HO states:
\begin{align}
  &\ket{N_\text{cm}L_\text{cm}, \,u_i\alpha;JM}\notag\\
  &\equiv\sum_{M_\text{cm} m}\braket{L_\text{cm}M_\text{cm}jm}{JM}\notag\\
  &\qquad\qquad\times\ket{N_\text{cm} L_\text{cm}M_\text{cm}} \otimes \ket{n\alpha jm}
  \label{eq:u_jacho_basis}\,,
\end{align}
where we have introduced the collective partial wave index $\alpha\equiv(lsTM_T)$ for brevity. The right singular vectors are constructed accordingly, and all singular vectors can be transformed separately by acting on them with the (unitary) Talmi-Moshinsky transformation matrix.

\begin{figure*}
  \setlength{\unitlength}{\textwidth}
  \begin{picture}(1.0000,0.3100)
    \put(0.0100,0.0000){\includegraphics[height=0.3100\unitlength,viewport=0 0 326 325,clip]{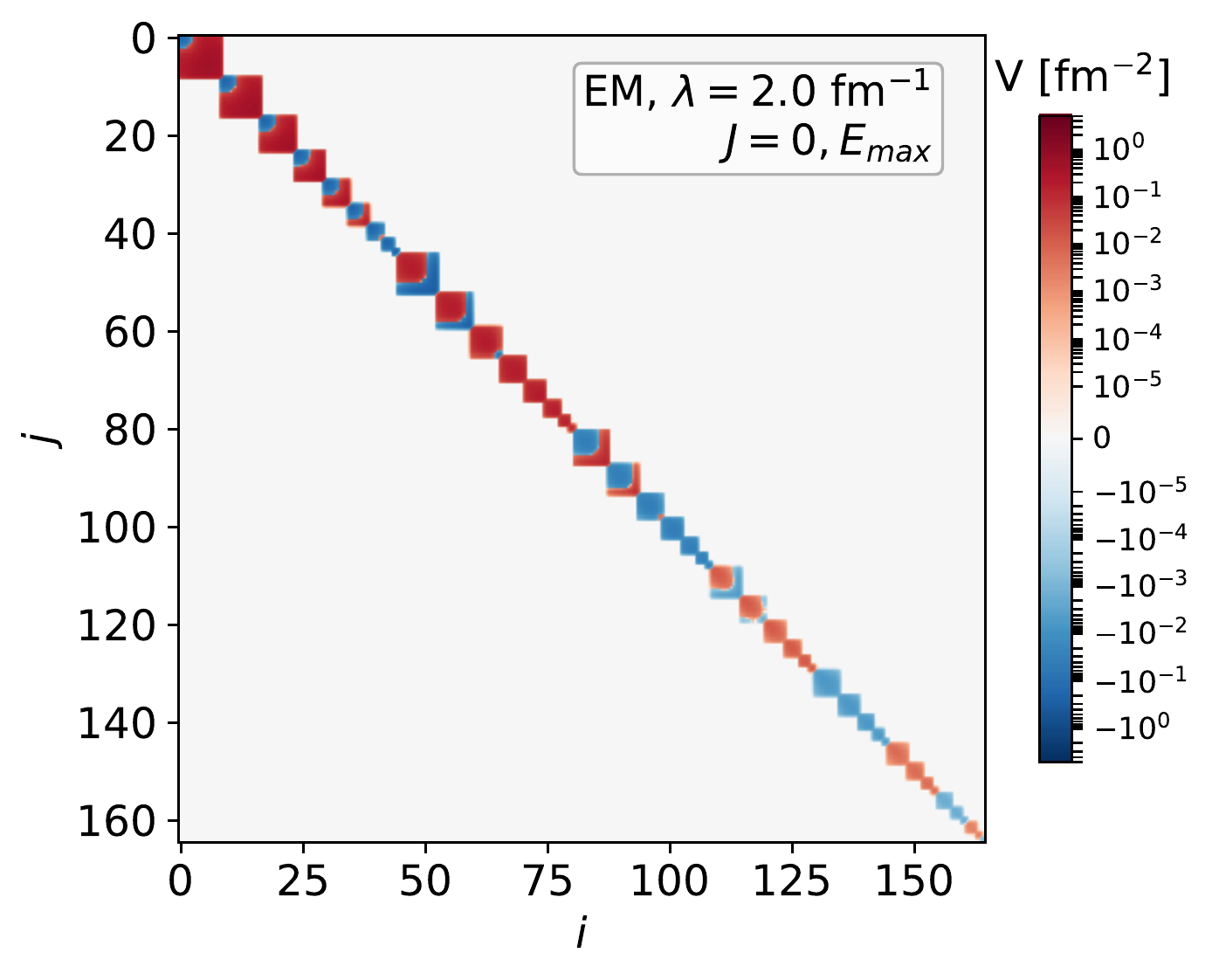}}
    \put(0.3333,0.0000){\includegraphics[height=0.3100\unitlength,viewport=22 0 326 325,clip]{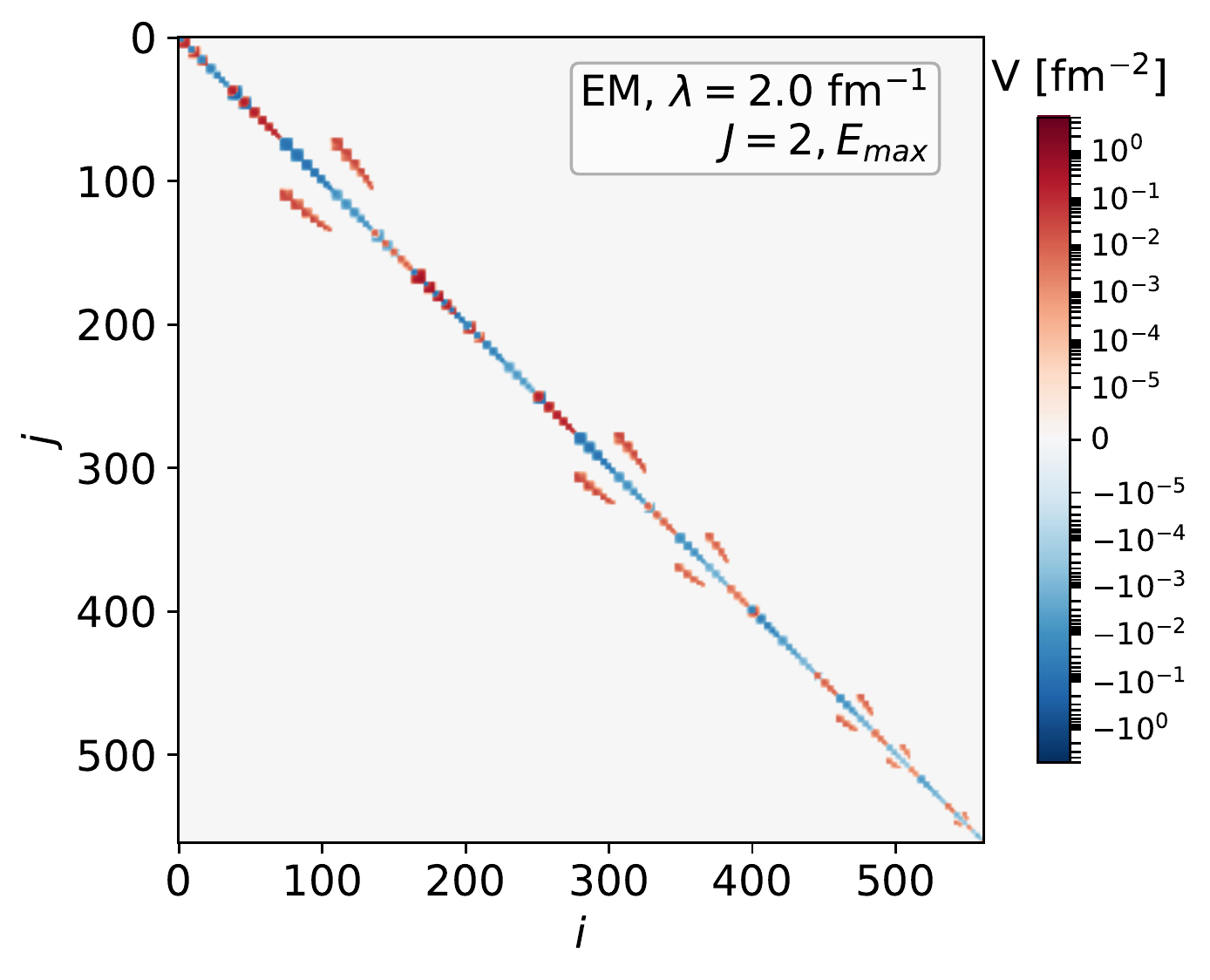}}
    \put(0.6366,0.0000){\includegraphics[height=0.3100\unitlength,viewport=22 0 400 325,clip]{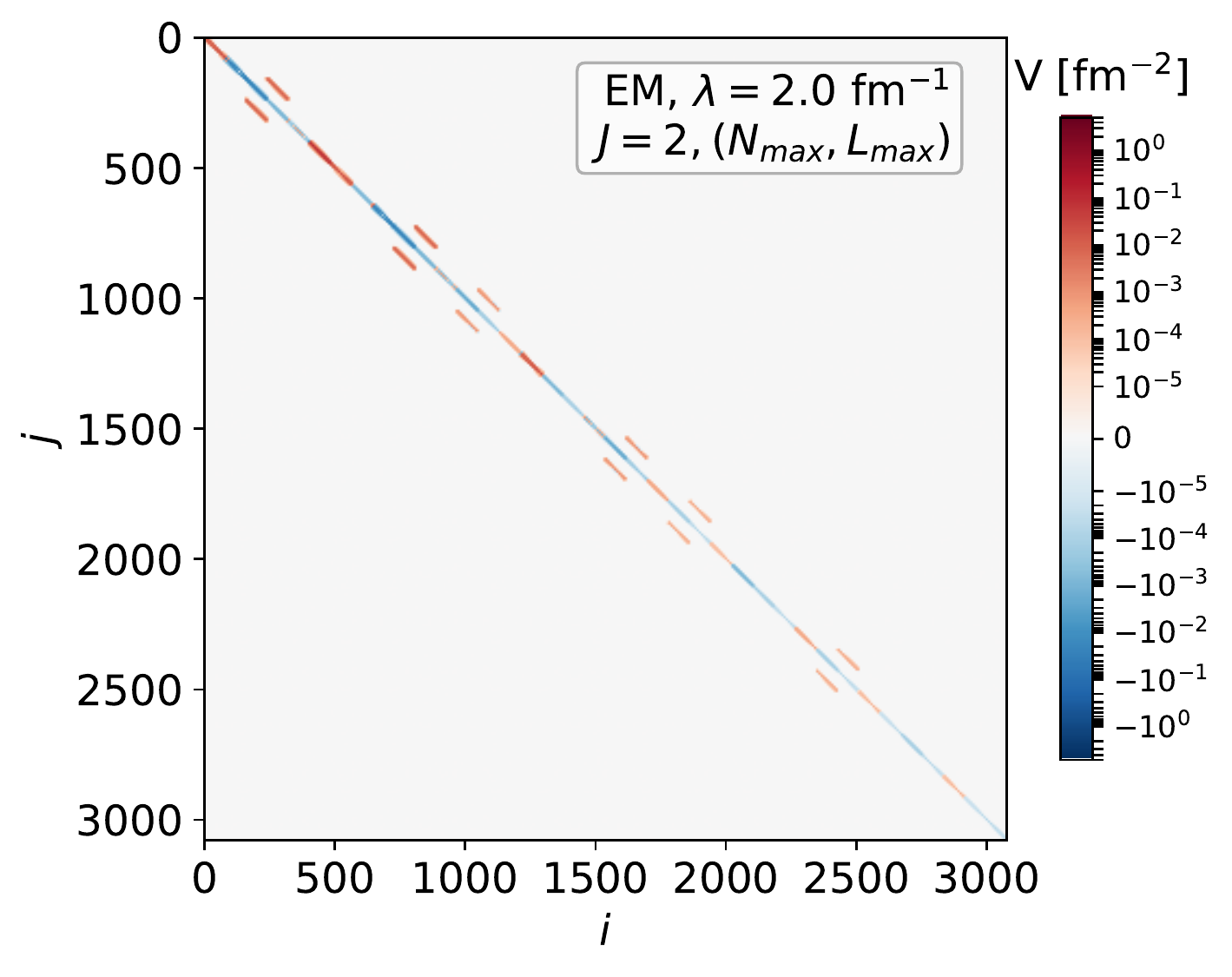}}
  \end{picture}
  \caption{
    Matrix elements in the $J=0$ and $J=2$ channels of the EM interaction at resolution scale $\lambda=2.0\,\fmi$,
    represented in the HO states $\ket{N_\text{cm}L_\text{cm},\,n(ls)j; JTM_T}$. The left and center panels are obtained with an $\EMax=16$ truncation, while the right panel uses $(\NMax,\LMax) = (8,16)$ (see text).
  }
  \label{fig:talmi}
\end{figure*}

It is clear from Eq.~\eqref{eq:u_jacho_basis} that each singular vector and singular value will be multiplied by the number of center-of-mass states. In this expanded basis, the matrix representation of $V$ is given by the Kronecker product of the identity matrix in the center-of-mass space with the factorized interaction in the relative space. As an example, we show the matrices obtained for the EM interaction at $\lambda=2.0\,\fmi$ in the $J,T,M_T=(0,1,0)$ and $(2,1,0)$ channels in Fig.~\ref{fig:talmi}. As we can see, the size and structure of the matrix depends on the truncation we impose on the oscillator basis: We can use the $\EMax$ truncation briefly discussed in Sec.~\ref{sec:ho}, which requires
\begin{align}
  E &= 2N_\text{cm} + L_\text{cm} + 2n+l \notag\\ &= 2n_1 + l_1 + 2n_2+l_2 \leq \EMax
\end{align}
(with single-particle oscillator quantum numbers $n_i, l_i$ in the laboratory frame), or we can introduce independent truncations $N_\text{cm}, n\leq \NMax$ and $L_\text{cm}, l\leq \LMax$. For the former, the size of the copies of the partial waves decreases as $N_\text{cm}$ (left and center panels of Fig.~\ref{fig:talmi}), and the size of the singular values changes due to the projection into the smaller space. For the latter, we obtain exact copies of the partial waves (right panel).

\begin{figure}
  \setlength{\unitlength}{\columnwidth}
  \begin{picture}(1.0000,1.02000)
     \put(0.0000,0.6400){\includegraphics[width=\columnwidth]{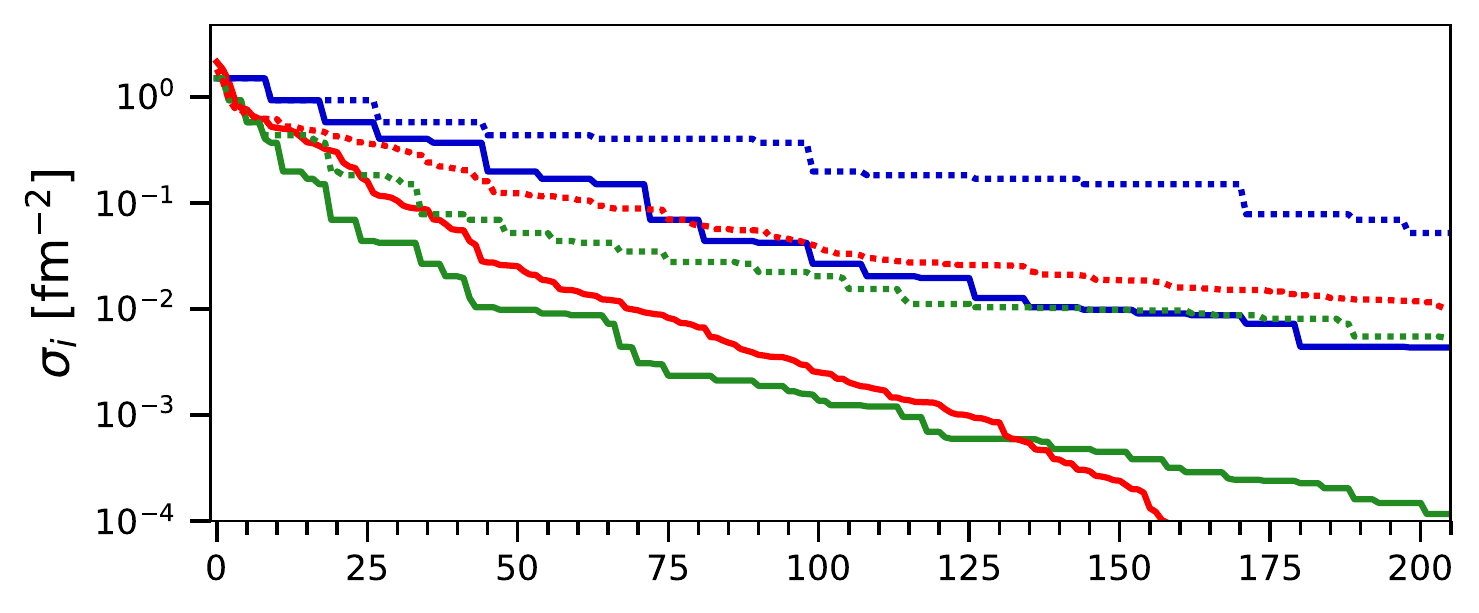}}
     \put(0.0000,0.0000){\includegraphics[width=\columnwidth]{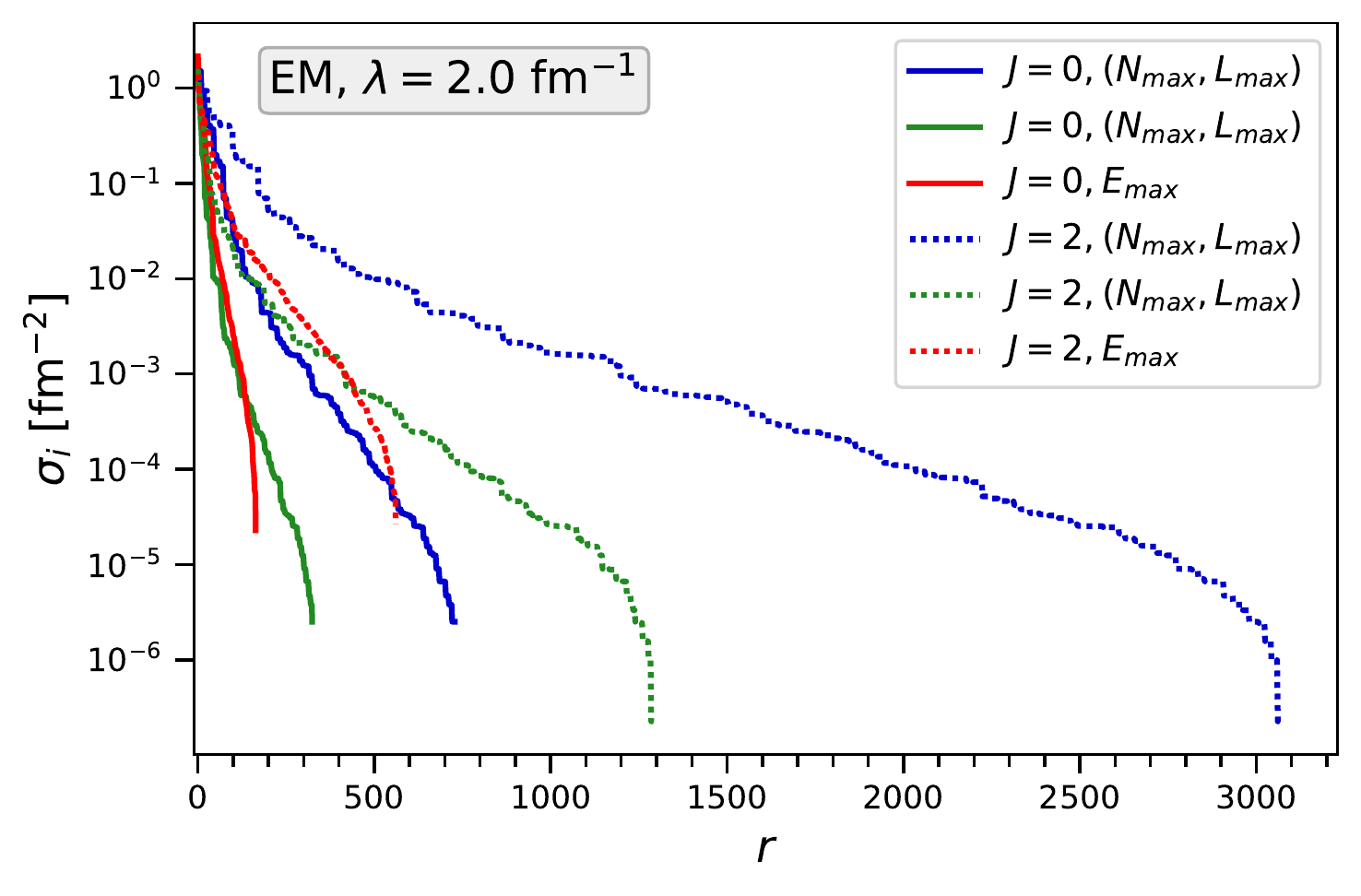}}
  \end{picture}
  \vspace{-20pt}
  \caption{
    Singular values of matrices shown in Fig.~\ref{fig:talmi} (plus the $J=0$ channel in $(\NMax,\LMax) = (8,16)$ 
    truncation). The green curves are obtained by compressing duplicate singular values from the full $(\NMax,\LMax) = (8,16)$
    sets.
  }
  \label{fig:talmi_svd}
\end{figure}

Figure \ref{fig:talmi_svd} shows the singular value spectra of these matrices. Unsurprisingly, the factorized matrix in $(\NMax,\LMax)$ truncation has many more relevant singular vectors than the $\EMax$ truncation, but it is readily compressible because we only need to store one representative for each group of identical copies of a given partial wave. At least formally, the factorized matrix in $\EMax$ truncation is not, because of the projection of the singular vectors into a lower-dimensional HO basis and the associated change of the singular values. While the ranks of the $\EMax$ and $(\NMax,\LMax)$ matrices are roughly similar overall, a detailed view of the dominant singular vectors gives the latter a slight advantage.

In the $(\NMax,\LMax)$ case, the rank of the interaction in each channel will be given by the sum of the ranks of the partial waves that can contribute to each channel under the usual selection rules for angular momentum and parity: For the $(J,T,M_T) = (0,1,0)$ channel, for example, the rank will be the sum of the ranks of all $T=1$ neutron-proton partial waves, since we can couple each relative angular momentum $j$ with the corresponding $L_\text{cm}$ to total angular momentum $J=0$. In the $(J,T,M_T) = (2,1,0)$, all partial waves with $|L_\text{cm}-2|\leq j \leq L_\text{cm} + 2$ are allowed, and this amount of allowed coupling will grow with the total $J$.  This matches the observations of a recent study that applied tensor factorization techniques to nuclear interactions, which found an increase of their rank with $J$ \cite{Tichai:2019xz}.

To conclude, we saw how the embedding of the factorized interactions into a larger space in the context of the Talmi-Moshinsky transformation introduces copies of the singular values that formally increase the rank of the interaction. Based on our analysis here, it seems most appropriate to tackle this issue by performing the transformation in $(\NMax, \LMax)$ truncation because then the copies will be identical and one easily avoid the additional storage. An amplified version of this issue appears in the implementation of the SVD-SRG for three-body forces, because the two-body relative partial waves must be embedded into the three-body relative partial waves to track induced forces \cite{Hebeler:2020ex}. Research on how to overcome this issue in the next stage of our project is in progress.

\section{Many-Body Calculations}
\label{sec:manybody}

\begin{figure}
\setlength{\unitlength}{\columnwidth}
\begin{picture}(1.0000,2.1000)
    \put(0.0100,1.4000){\centering\includegraphics[width=0.96\unitlength]{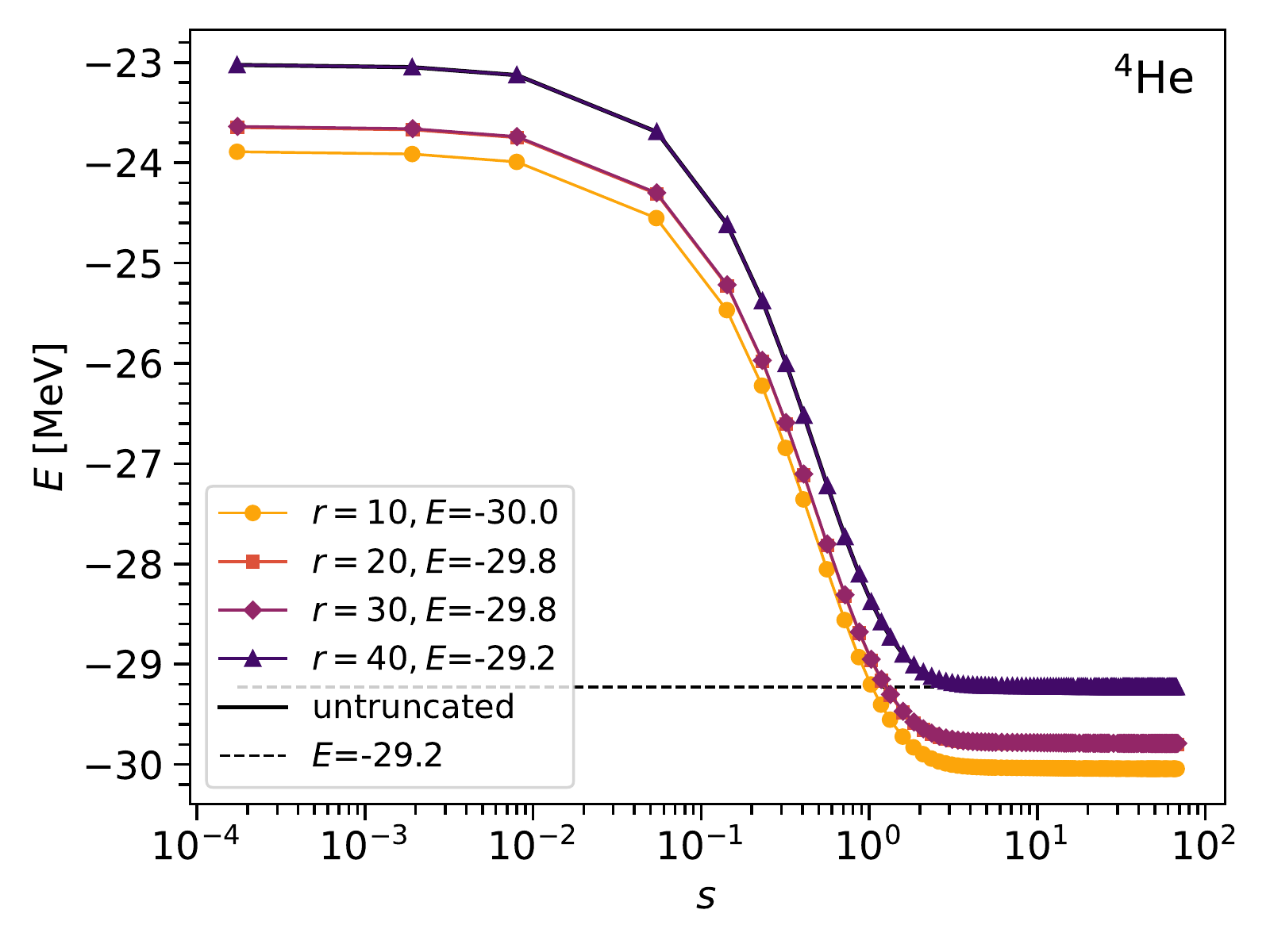}}
    \put(0.0000,0.7000){\centering\includegraphics[width=0.96\unitlength]{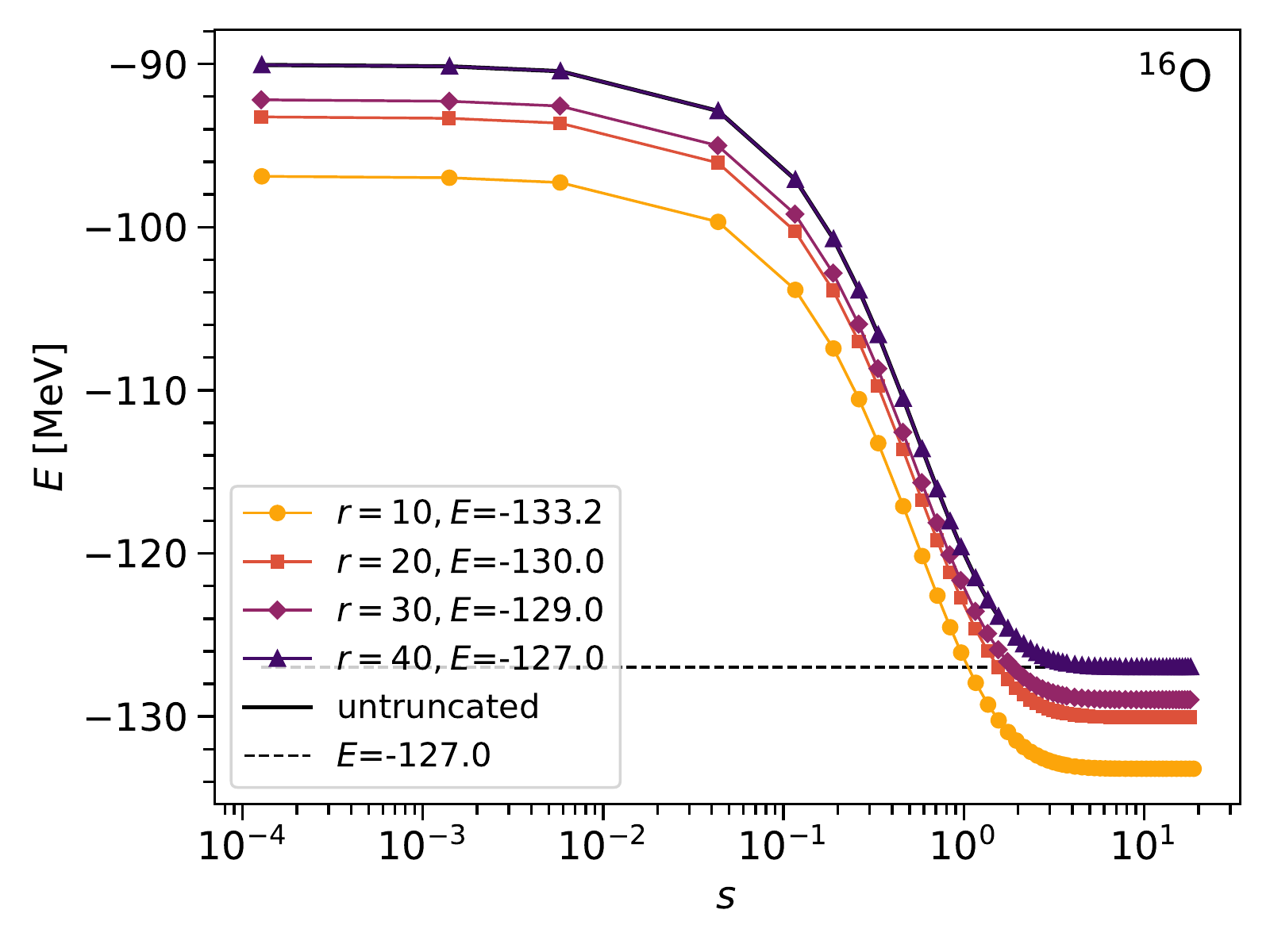}}
    \put(0.0000,0.0000){\centering\includegraphics[width=0.96\unitlength]{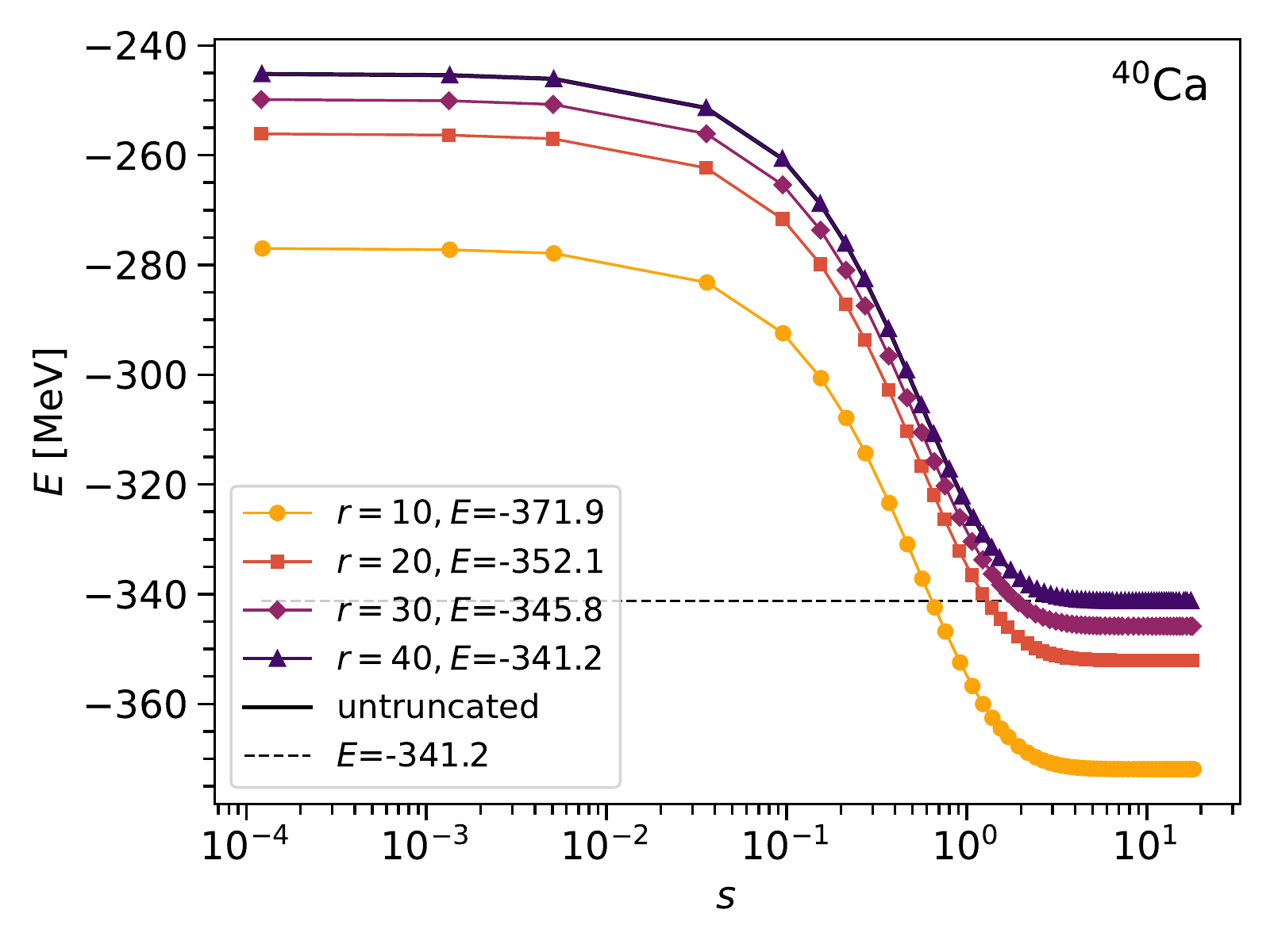}}
\end{picture}
\caption{IMSRG(2) ground-state energies of selected closed-shell nuclei as a function of the flow parameter $s$ for the EM1.8/2.0 $NN+3N$ interaction (see \cite{Hebeler:2011dq,Nogga:2004il} and text). The SVD-SRG with different ranks (per partial wave) is used to construct the evolved $NN$ component of the interaction. The results are obtained for a HO basis with $\eMax=8$ and $\EMax[3]=14$, which is sufficiently close to convergence in these nuclei.}
\label{fig:imsrg}
\end{figure}

After our extensive discussion of how the SVD and SVD-SRG can be integrated into the typical workflows for processing nuclear $NN$ and $3N$ interactions, we have now reached the final stage, applications in actual many-body calculations. Since it will be a formidable task to re-formulate current many-body methods to leverage the factorization for efficiency gains, we focus for now on benchmarking the accuracy of the rank-$r$ SVDs by performing conventional many-body calculations with the reconstructed interactions. 

In Fig.~\ref{fig:imsrg}, we show the results from ground-state energy calculations for closed-shell nuclei in the IMSRG(2) approach \cite{Hergert:2016jk,Hergert:2017kx}. They are generated using the so-called EM1.8/2.0, which consists of the EM interaction evolved to $\lambda=1.8\,\fmi$ and an \NNLO{} $3N$ interaction with cutoff $\Lambda=2.0\,\fmi$ whose low-energy constants have been fitted to the triton binding energy and $\nuc{He}{4}$ charge radius \cite{Nogga:2004il,Hebeler:2011dq}. While not fully consistent from the view of chiral EFT, this interaction has been empirically successful for the description of ground-state energies of a wide range of nuclei, although it underestimates radii by a few percent (see \cite{Hergert:2020am} and references therein, in particular \cite{Stroberg:2021qu}). It serves as a ``realistic'' complement to benchmark calculations that are based on the SVD-SRG evolved $NN$ interaction alone, which produce nuclei that are overbound and much too small. The performance of the rank$-r$ approximation and SVD-SRG is effectively the same in all the cases we studied.

The SVD-SRG interaction accurately recovers the results obtained without factorization once we include between 30 and 40 components per partial wave, which is consistent with our findings in the two-nucleon system. This encompasses the Hartree-Fock calculation that is used to prepare the reference state \cite{Hergert:2016jk,Hergert:2017kx}, as well as the details of the IMSRG(2) flow of the ground-state energy as a function of the flow parameter $s$. As before, this rank is primarily determined by the SVD of the Coulomb interaction $V_C$ between the protons, while 5 to 10 components provide a highly accurate reproduction if only nuclear interactions are included in the calculation.

\begin{figure}
  \includegraphics[width=0.48\textwidth]{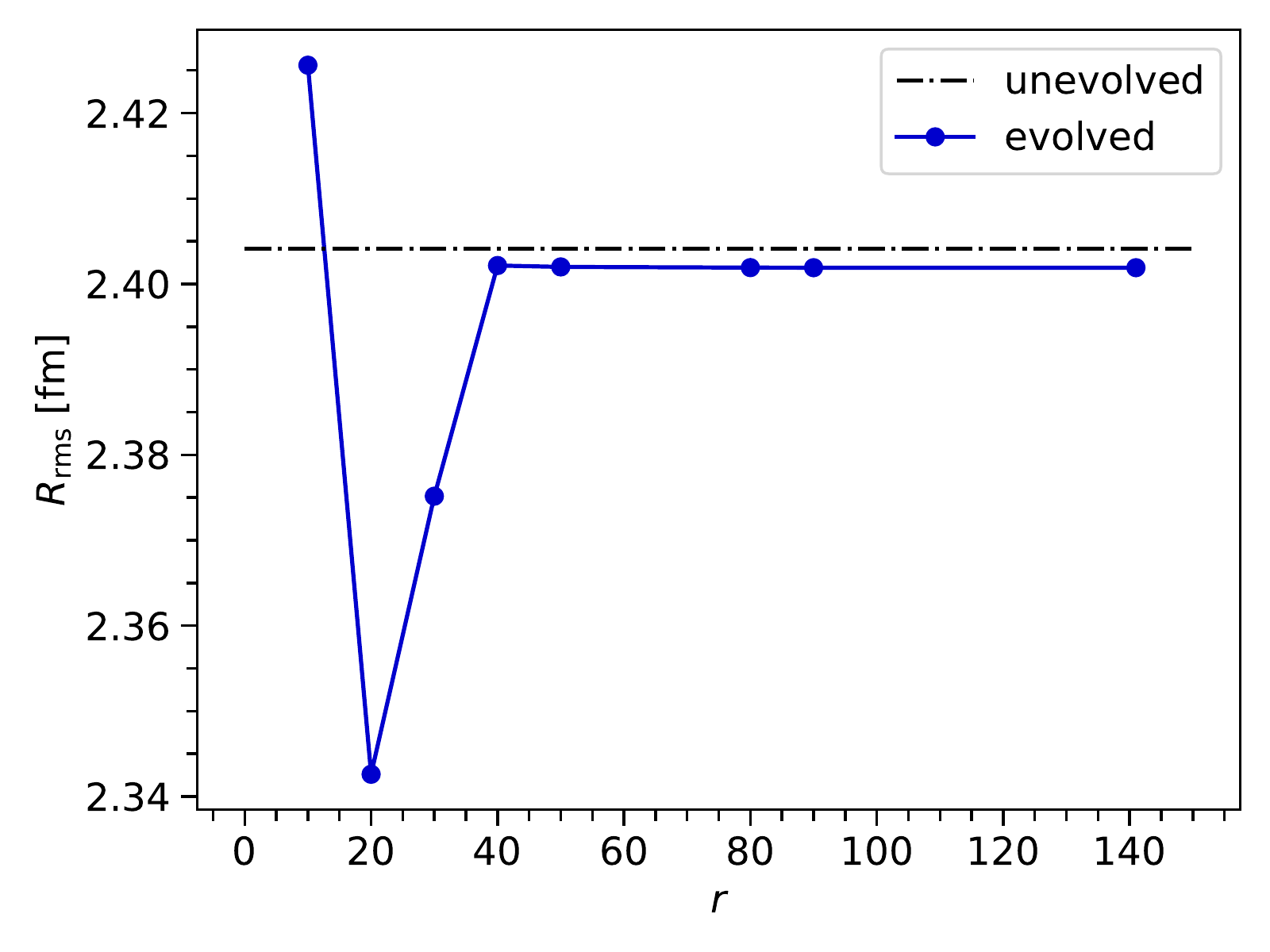}
  \vspace{-20pt}
  \caption{Root-mean-square radius of $^{40}\text{Ca}$ for the EM interaction at SRG resolution scale $\lambda=1.8\,\fm^{-1}$, using the evolved and unevolved operators.} 
  \label{fig:radius}
\end{figure}

To investigate the SVD-SRG evolution of general observables, we also constructed the the mean-square radius operator 
\begin{equation}
    R^2_{\text{ms}} = \frac{1}{A^2}\sum_{i=1}^A(r_i-R_{\text{cm}})^2 
\end{equation}
using Eq.~\eqref{eq:unitarySVD} to obtain the unitary transformation in the two-body system from the singular vectors of the truncated SVD. In Fig.~\ref{fig:radius}, we illustrate the dependence of the IMSRG(2) root-mean-square radius $R_\text{rms}\equiv\sqrt{\expect{R^2_\text{ms}}}$ of $\nuc{Ca}{40}$ on the rank of the SVD. As for the energy, 30 to 40 components are sufficient for an accurate reconstruction of the unitary transformation. For comparison, we also include the result obtained with the unevolved $R^2_\text{ms}$ operator, which is larger by about $0.01\,\fm$. Since the SRG evolution targets physics at high-momentum or short-range, its effect on a long-ranged operator like $R^2_\text{ms}$ is weak, and negligible compared to other sources of uncertainty at present. We conclude our discussion by remarking that while $R^2_\text{ms}$ is certainly one of the simplest operators besides the energy that can be investigated, we do not expect issues in applying the factorized unitary transformation to more complex operators, e.g., in studies of electroweak transitions \cite{Parzuchowski:2017ta,Gysbers:2019df,Yao:2020mw}, since it is completely determined by the properties of the Hamiltonian.

\section{Conclusions and Outlook}

In this work, we have used the Singular Value Decomposition (SVD) to performed principal-component analyses of two current nucleon-nucleon interactions, the frequently used chiral \NNNLO{} interaction by Entem and Machleidt \cite{Entem:2003th} and the Argonne V18 interaction \cite{Wiringa:1995or}. We showed that the former readily allows the construction of a low-rank representation by truncating the SVD based on the size of the singular values, while the situation is much more complicated for the latter because of its local nature and the ensuing structure in momentum space, and its high implicit resolution scale.

We have merged the SVD with Similarity Renormalization Group (SRG) techniques, and shown that the factorized representation can be accurately evolved to lower resolution scales, although the rank of the initial interaction ultimately determines whether it is significantly more efficient than the traditional SRG. While the SRG evolution of two-nucleon interactions is no challenge nowadays, we intend to extend these techniques to three-nucleon forces in the next stage of our project, where more significant efficiency gains are possible. Looking even further into the future, an additional extension to the four-body system might make the consistent evolution of initial \cite{Epelbaum:2006ij,Rozpedzik:2006lv,Kaiser:2012xy} and induced four-nucleon forces possible \cite{Calci:2014xy,Schulz:2018vp,Hebeler:2020ex}.  

In the present work, we have carried the factorized form of the nucleon-nucleon interactions through the major steps of the workflows that are used to prepare them for application in nuclear many-body calculation, but the major task to formulate current many-body methods themselves to exploit the factorization remains for the future. In combination with an SVD-SRG for three-nucleon forces, it holds the potential for unparalleled efficiency gains both in the storage requirements and the computational cost. This is essential as calculations are pushed to heavier, more exotic, and structurally more complex nuclei.

While these developments are the main focus of our own efforts, the present study suggests additional directions for future research. As we have seen throughout our discussion, the consistent inclusion of the Coulomb interaction in the SVD-SRG evolution adversely impacts the rank compression that can be achieved in proton-proton channels of the interaction, hence it is worth our while to explore alternative treatments, e.g., by handling it similarly as the kinetic energy. Next, the remarkable similarity of the interaction's rank across all partial waves could suggests that we are merely seeing the projection of a few relevant operators into different channels, hence we will explore whether such a connection can indeed be made. While the momentum and HO bases explored for the SVD-SRG in this work have somewhat similar characteristics, the observation that we obtain nearly identical ranks both bases may be further evidence in favor of this hypothesis. Finally, the implementation of the Talmi-Moshinsky transformation to the SVD factors was a prototypical example for the artificial increase of how the embedding of an operator in larger product Hilbert spaces increases the rank of the operator's matrix representations through redundant copies. Given that the many-body Hilbert spaces themselves have a product structure, tensor representations seem like a particularly suitable candidate for handling the physical information encoded in an operator in the most efficient way --- indeed, this is the reason for the success of tensor network methods in other areas of many-body physics.

\textbf{Note:} While this work was in its late stages, a preprint about low-rank decompositions of chiral nucleon-nucleon interactions was published \cite{Tichai:2021cd}. This study some overlap with our work and provides independent confirmation for part of the results presented here.

\section*{Acknowledgments}
We thank C. Haselby, M. Iwen, J. M. Yao, and A. Zare for useful discussions. 

This work has been supported by the U.S. Department of Energy, Office of Science, 
Office of Nuclear Physics under Awards No. de-sc0017887 and No. de-sc0018083 
(NUCLEI SciDAC-4 Collaboration).

\bibliography{references,preprints}
\end{document}